\documentclass[dvips]{article}
\usepackage{supertabular,lscape,epsfig}
\usepackage{amssymb}
\usepackage{amsmath}
\usepackage{rotating}
\DeclareSymbolFont{ppa}{OT1}{ppl}{m}{it}
\DeclareMathSymbol{\vv}{\mathalpha}{ppa}{'166}

\begin{document}

\newcommand{\dd}{\,{\rm d}}
\newcommand{\ie}{{\it i.e.},\,}
\newcommand{\etal}{{\it et al.\ }}
\newcommand{\eg}{{\it e.g.},\,}
\newcommand{\cf}{{\it cf.\ }}
\newcommand{\vs}{{\it vs.\ }}
\newcommand{\zdot}{\makebox[0pt][l]{.}}
\newcommand{\up}[1]{\ifmmode^{\rm #1}\else$^{\rm #1}$\fi}
\newcommand{\dn}[1]{\ifmmode_{\rm #1}\else$_{\rm #1}$\fi}
\newcommand{\upd}{\up{d}}
\newcommand{\uph}{\up{h}}
\newcommand{\upm}{\up{m}}
\newcommand{\ups}{\up{s}}
\newcommand{\arcd}{\ifmmode^{\circ}\else$^{\circ}$\fi}
\newcommand{\arcm}{\ifmmode{'}\else$'$\fi}
\newcommand{\arcs}{\ifmmode{''}\else$''$\fi}
\newcommand{\MS}{{\rm M}\ifmmode_{\odot}\else$_{\odot}$\fi}
\newcommand{\RS}{{\rm R}\ifmmode_{\odot}\else$_{\odot}$\fi}
\newcommand{\LS}{{\rm L}\ifmmode_{\odot}\else$_{\odot}$\fi}
\newcommand{\feh}{\hbox{$ [{\rm Fe}/{\rm H}]$}}

\newcommand{\Abstract}[2]{{\footnotesize\begin{center}ABSTRACT\end{center}
\vspace{1mm}\par#1\par
\noindent
{~}{\it #2}}}

\newcommand{\TabCap}[2]{\begin{center}\parbox[t]{#1}{\begin{center}
  \small {\spaceskip 2pt plus 1pt minus 1pt T a b l e}
  \refstepcounter{table}\thetable \\[2mm]
  \footnotesize #2 \end{center}}\end{center}}

\newcommand{\TableSep}[2]{\begin{table}[p]\vspace{#1}
\TabCap{#2}\end{table}}

\newcommand{\FigCap}[1]{\footnotesize\par\noindent Fig.\  %                     
  \refstepcounter{figure}\thefigure. #1\par}

\newcommand{\TableFont}{\footnotesize}
\newcommand{\TableFontIt}{\ttit}
\newcommand{\SetTableFont}[1]{\renewcommand{\TableFont}{#1}}

\newcommand{\MakeTable}[4]{\begin{table}[htb]\TabCap{#2}{#3}
  \begin{center} \TableFont \begin{tabular}{#1} #4
  \end{tabular}\end{center}\end{table}}

\newcommand{\MakeTableSep}[4]{\begin{table}[p]\TabCap{#2}{#3}
  \begin{center} \TableFont \begin{tabular}{#1} #4
  \end{tabular}\end{center}\end{table}}

\newenvironment{references}%                                                    
{
\footnotesize \frenchspacing
\renewcommand{\thesection}{}
\renewcommand{\in}{{\rm in }}
\renewcommand{\AA}{Astron.\ Astrophys.}
\newcommand{\AAS}{Astron.~Astrophys.~Suppl.~Ser.}
\newcommand{\ApJ}{Astrophys.\ J.}
\newcommand{\ApJS}{Astrophys.\ J.~Suppl.~Ser.}
\newcommand{\ApJL}{Astrophys.\ J.~Letters}
\newcommand{\AJ}{Astron.\ J.}
\newcommand{\IBVS}{IBVS}
\newcommand{\PASJ}{PASJ}
\newcommand{\PASP}{P.A.S.P.}
\newcommand{\Acta}{Acta Astron.}
\newcommand{\MNRAS}{MNRAS}
\renewcommand{\and}{{\rm and }}
\section{{\rm REFERENCES}}
\sloppy \hyphenpenalty10000
\begin{list}{}{\leftmargin1cm\listparindent-1cm
\itemindent\listparindent\parsep0pt\itemsep0pt}}%                               
{\end{list}\vspace{2mm}}

\def\TYLDA{~}
\newlength{\DW}
\settowidth{\DW}{0}
\newcommand{\dw}{\hspace{\DW}}

\newcommand{\refitem}[5]{\item[]{#1} #2%                                        
\def\REFARG{#3}\ifx\REFARG\TYLDA\else, {\it#3}\fi
\def\REFARG{#4}\ifx\REFARG\TYLDA\else, {\bf#4}\fi
\def\REFARG{#5}\ifx\REFARG\TYLDA\else, {#5}\fi.}

\newcommand{\Section}[1]{\section{#1}}
\newcommand{\Subsection}[1]{\subsection{#1}}
\newcommand{\Acknow}[1]{\par\vspace{5mm}{\bf Acknowledgments.} #1}
\pagestyle{myheadings}

\newfont{\bb}{ptmbi8t at 12pt}
\newcommand{\xrule}{\rule{0pt}{2.5ex}}
\newcommand{\xxrule}{\rule[-1.8ex]{0pt}{4.5ex}}
\def\thefootnote{\fnsymbol{footnote}}

\begin{center}
{\Large\bf Over 10~000 Delta Scuti Stars toward the Galactic Bulge from
OGLE-IV\footnote{Based on observations obtained with the 1.3-m Warsaw telescope
at the Las Campanas Observatory of the Carnegie Institution for Science.}}
\vskip1cm
{\bf
P.~~P~i~e~t~r~u~k~o~w~i~c~z$^1$,~~I.~~S~o~s~z~y~\'n~s~k~i$^1$,~~H.~~N~e~t~z~e~l$^2$,\\
~~M.~~W~r~o~n~a$^1$,~~A.~~U~d~a~l~s~k~i$^1$,~~M.~K.~~S~z~y~m~a~\'n~s~k~i$^1$,\\
~~R.~~P~o~l~e~s~k~i$^1$,~~S.~~K~o~z~{\l}~o~w~s~k~i$^1$,~~J.~~S~k~o~w~r~o~n$^1$,\\
~~K.~~U~l~a~c~z~y~k$^{1,3}$,~~D.~M.~~S~k~o~w~r~o~n$^1$,~~P.~~M~r~\'o~z$^{1,4}$,\\
~~K.~~R~y~b~i~c~k~i$^1$,~~P.~~I~w~a~n~e~k$^1$,~~and~~M.~~G~r~o~m~a~d~z~k~i$^1$\\}
\vskip3mm
{
$^1$ Astronomical Observatory, University of Warsaw, Al. Ujazdowskie 4, 00-478 Warszawa, Poland\\
$^2$ Nicolaus Copernicus Astronomical Center, ul. Bartycka 18, 00-716 Warszawa, Poland\\
$^3$ Department of Physics, University of Warwick, Coventry CV4 7AL, UK\\
$^4$ Division of Physics, Mathematics, and Astronomy, California Institute of Technology, Pasadena, CA 91125, USA\\
}
\end{center}

\Abstract{We present a collection of 10~111 genuine $\delta$~Sct-type pulsating
variable stars detected in the OGLE-IV Galactic bulge fields. In this sample,
9835 variables are new discoveries. For most of the stars photometric data
cover the whole decade 2010--2019. We illustrate a huge variety of light
curve shapes of $\delta$~Sct variables. Long-term observations have allowed
us to spot objects with evident period, amplitude, and mean brightness
variations. Our analysis indicates that about 28\% of the stars are
single-mode pulsators. Fourteen $\delta$~Sct stars show additional eclipsing
or ellipsoidal binary modulation. We report significant attenuation or
even disappearance of the pulsation signal in six sources. The whole set
of variables is a mix of objects representing various Milky Way's
populations, with the majority of stars from the Galactic bulge.
There are also representatives of the Sagittarius Dwarf Spheroidal Galaxy.
Some of the newly detected variables could be SX Phe-type stars residing in
globular clusters. The collection, including full $V$- and $I$-band time-series data,
is available to the astronomical community from the OGLE On-line Data Archive.}

{Catalogs -- Stars: variables: $\delta$~Scuti -- Stars: variables: SX~Phoenicis
-- Galaxy: bulge -- Galaxies: individual: Sagittarius Dwarf Spheroidal Galaxy}

%%%%%%%%%%%%%%%%%%%%%%%%%%%%%%%%%%%%%%%%%%%%%%%%%%%%%%%%%%%%%%%%%%%%%%%%%%%

\Section{Introduction}

$\delta$~Sct-type variables are pulsating stars with periods below 0.3~d
and $V$-band amplitudes up to 0.9~mag. They pulsate in radial as well as
non-radial acoustic modes excited mainly in the $\kappa$ mechanism
(Breger 2000). The majority of $\delta$~Sct stars are multiple-mode pulsators.
The spectral types of $\delta$~Sct variables range from A0 to F6 for
luminosity classes III (giants), IV (subgiants), and V (dwarfs).
In the Hertzsprung-Russell diagram, these pulsating stars lie at the
classical instability strip on the main sequence (MS) or are moving from
the MS to the giant branch. They can also be found at the pre-MS stage.
The stars belong to various populations. Usually, $\delta$~Sct stars are
considered the Population~I stars of the flat Milky Way component (the
young Galactic disk). Population~II analogues or representatives of the
Galactic halo and old disk are sometimes classified as SX Phe stars.
This type of variables is observed in globular clusters. SX Phe stars
are common among blue stragglers in metal-poor clusters, but relatively
rare in metal-rich ones. The mass range of $\delta$ Sct stars
depends on the metal content. Population~I pulsators cover
a range from about 1.5 to 2.3~$\MS$ (Murphy \etal 2019),
while Population~II stars from about 1.0 to 1.3~$\MS$ (McNamara 2011).
Space-based observations have shown that vast majority of $\delta$~Sct
stars are low-amplitude pulsators ($<0.1$~mag) and only about half of
them have amplitudes $>0.001$~mag and appear to be variable from the
ground (Balona and Dziembowski 2011).

The prototype object of the whole class, $\delta$~Sct itself,
was recognized as a star exhibiting radial velocity variations
amongst nine 4th magnitude stars observed by Campbell and Wright (1900).
It was thought to be a spectroscopic binary system. Much later,
results from photometric and spectroscopic observations of this object
reported by Fath (1935) and Colacevich (1935), respectively, showed
a short variability period of 0.1937~d and small amplitude of the
velocity variations of about 8 km/s. The relation between the radial
velocity changes and light variations (nearly mirrored phased curves)
was identical to those observed in Cepheids. After the discovery of
multiple variability in $\delta$~Sct (Fath 1937, Sterne 1938),
it was natural to interpret the behavior of the star in terms of
the pulsation theory. So far, six distinct modes in the power
spectrum of $\delta$~Sct have been detected (Templeton \etal 1997).

By December 1956, or within two decades from the identification
of $\delta$~Sct as the pulsating star, four variables of this type were
discovered: DQ Cep (Walker 1952), CC And (Lindblad and Eggen 1953),
$\rho$~Pup (Eggen 1956a), and $\delta$~Del (Eggen 1956b). Shortly
thereafter, the number of new $\delta$~Sct stars (often termed dwarf
Cepheids in the past) started to increase much faster and reached 636
sources in January 2000, in the catalog prepared by Rodr\'iguez \etal (2000).

A truly rapid increase began in the 1990s with the advent of CCD detectors
and wide-field photometric surveys. Dozens of $\delta$~Sct stars were found
as by-products from microlensing surveys. The Optical Gravitational Lensing
Experiment during its first phase of operation (OGLE-I, years 1992--1995)
discovered 53 $\delta$~Sct stars mainly in the area of Baade's Window
of the Galactic bulge (Udalski \etal 1994, 1995ab, 1996, 1997).
The MAssive Compact Halo Object (MACHO) project reported on the detection
of 90 $\delta$~Sct stars toward the bulge, 86 of which were new variables
(Alcock \etal 2000). Extensive time-series observations of Galactic
globular clusters led to the discovery of tens of SX Phe-type stars
(\eg Kaluzny \etal 1996, Pych \etal 2001, Mazur \etal 2003). In their catalog,
Rodr\'iguez and L\'opez-Gonz\'alez (2000) list 149 such variables
belonging to eighteen globular clusters and two nearby galaxies (Carina
and Sagittarius Dwarf Spheroidal Galaxies). Pigulski \etal (2006)
analyzed OGLE-II photometry (collected in years 1997--2000) and reported
on the detection of 193 high-amplitude $\delta$~Sct stars, including
50 multi-periodic objects. In OGLE-III data (years 2001--2009)
for the Large Magellanic Cloud, Poleski \etal (2010) found 2786
short-period variables, 92 of which were multi-mode pulsators.

Large-scale surveys have multiplied the number of known Galactic $\delta$~Sct
stars. The All-Sky Automated Survey (ASAS; Pojma\'nski 2002) led to the
detection of 525 previously unknown $\delta$~Sct objects. The fifth edition
of the General Catalogue of Variable Stars (GCVS; Samus \etal 2017, version
from November 2020) contains 1019 positions classified as $\delta$~Sct
(DSCT) and 228 positions classified as SX Phe stars (SXPHE). A compiled
catalog of 1578 Galactic $\delta$~Sct pulsators was presented by
Chang \etal (2013). There were 4514 DSCT and 7 SXPHE identifications in the
International Variable Star Index (VSX; Watson \etal 2006) in May 2019.
Very recently, Jayasinghe \etal (2020) presented an all-sky catalog of
8418 $\delta$~Sct variables from the All-Sky Automated Survey for SuperNovae
(ASAS-SN; Shappee \etal 2014, Kochanek \etal 2017). According to the
authors, the catalog includes 3322 new discoveries. Finally,
Chen \etal (2020) published a set of periodic variable stars
containing 15~396 candidate $\delta$~Sct pulsators detected
mostly in the northern hemisphere ($\delta >-25\arcd$)
by the Zwicky Transient Facility (ZTF; Bellm \etal 2019).

$\delta$~Sct stars were intensively observed by Kepler space
telescope. A thorough search for such variables was recently carried out by
Murphy \etal (2019). They identified 1988 genuine $\delta$~Sct pulsators
in the original 105 deg$^2$ {\it Kepler} field monitored for four years
almost continuously (from May 2009 to May 2013). Currently, among
millions of stellar sources, $\delta$~Sct stars are observed by Gaia
and TESS space missions.

Here, we introduce a collection of 10~111 genuine $\delta$~Sct variable
stars detected in the OGLE-IV Galactic bulge fields. We cross-match
our collection with previously published catalogs, including lists
of variable stars in globular clusters. In the entire sample, 9835 objects
are newly discovered variables. This release significantly increases
the number of known $\delta$~Sct stars in the whole sky. In the paper,
we present some general observational properties of the variables
and a huge diversity of their light curves.

%%%%%%%%%%%%%%%%%%%%%%%%%%%%%%%%%%%%%%%%%%%%%%%%%%%%%%%%%%%%%%%%%%

\Section{OGLE Observations and Data Reductions}

The photometric data used in this work were collected with the 1.3-m
Warsaw telescope at Las Campanas Observatory (LCO), Chile, during the
fourth phase of the OGLE project (OGLE-IV) in years 2010--2019.
LCO is operated by the Carnegie Institution for Science.
The OGLE-IV mosaic camera consists of 32 2K$\times$4K CCD detectors
covering a total field of view of 1.4~deg$^2$ at a scale of 0.26 arcsec/pixel.
OGLE monitors optical variability in the area of the Galactic bulge,
Galactic disk, and Magellanic System. A total of about 710~000 exposures were
collected over the mentioned decade. Around 94\% of the frames were taken
through the Cousins $I$ filter. The remaining 6\% of the frames were taken 
through the Johnson $V$ filter. Details on the instrumentation setup can be
found in Udalski \etal (2015).

The presented collection of $\delta$~Sct stars is based on observations
of 121 OGLE-IV Galactic bulge fields covering an area of approximately
172~deg$^2$. Seven of the fields are located in the central part of the
tidally disrupted Sagittarius Dwarf Spheroidal Galaxy (Sgr dSph). A total
number of about 181~500 $I$-band exposures with the integration time of 100~s
(150~s in the case of Sgr dSph) and 6400 $V$-band exposures with 150~s
were obtained. The fields were monitored with various cadence depending
mainly on the frequency of microlensing events (Udalski \etal 2015).
In most crowded fields, such as BLG501, BLG505, and BLG512, one exposure
was taken every 19~min. In general, the number of $I$-band measurements
varies from 61 up to 16~799 per field, with a median value of 606.
In the $V$ band, there are from several up to 230 observations.
Some of the fields were not observed for the whole decade.
Least crowded fields, including fields close to the Galactic plane,
were monitored in the first 1--4 seasons only. The investigated bulge area
contains a total number of $\approx4.0 \times 10^8$ stars with brightness
between $I\approx12.5$~mag and $I\approx21.5$~mag. The photometry was obtained
with the standard OGLE data reduction pipeline using the Difference
Image Analysis (DIA; Alard and Lupton 1998, Wo\'zniak 2000).

In the first step, we carried out a frequency search up to 100~d$^{-1}$
with the help of the F{\footnotesize NPEAKS}
code\footnote{http://helas.astro.uni.wroc.pl/deliverables.php?lang=en\&active=fnpeaks}
for all $I$-band light curves from seasons 2010--2013. It calculates
Fourier amplitude spectra of unequally spaced time-series data. The code
reduces the computation time for a discrete Fourier transform by co-adding
correctly phased, low-resolution Fourier transforms of pieces of the large
data set interpolated to high resolution.

In the next step, light curves with high variability signal-to-noise
ratio (over 10) were visually inspected. Objects with asymmetric light curves
and periods shorter than 0.3~d, excluding evident exceptions, were classified
as $\delta$~Sct candidates. Based on our long-term experience in the
field of variable stars, we were able to recognize and separate from
the sample objects such as RRc variables (Soszy\'nski \etal 2014),
blue large-amplitude pulsators (BLAPs; Pietrukowicz \etal 2017),
eclipsing and ellipsoidal contact binary systems (Soszy\'nski \etal 2016).

The initially selected sample required a removal of outlying points
from the light curves. For objects with less than 1000 $I$-band measurements
the outliers were removed manually. In the case of objects with more
observations, which constitute about 57\% of all stars, we applied a
$3\sigma$-clipping procedure to phased light curves. After cleaning the data,
we improved the periods and re-inspected the whole sample visually.
The most interesting examples of $\delta$~Sct light curves were selected
and they are presented in Section~5. The period values were corrected
with the T{\footnotesize ATRY} code (Schwarzenberg-Czerny 1996) based
on the entire available time span (2010--2019). This code employs periodic
orthogonal polynomials to fit the data and the analysis of variance
(ANOVA) statistic to evaluate the quality of the fit.

During the inspection, we noticed that many $\delta$~Sct light curves
show a scatter characteristic for multi-mode pulsators. We performed
a pre-whitening of the data and we looked for secondary periods.
We found 2880 single-mode pulsators. This constitutes about 28\% of
the whole sample. Searching for and detailed analysis of additional
periodicities is the topic of the work by Netzel \etal (2021, in prep.).
Here, we present several examples of double-mode $\delta$~Sct stars and
candidates for members of binary systems.

Finally, each light curve was calibrated from the instrumental to the standard
magnitudes according to the prescription given in Udalski \etal (2015).
The accuracy of the calibration reaches 0.02 mag in the Johnson-Cousins
system. For about 5\% of the variables, located mainly in highly-extincted
regions, there is no $V$-band measurement. In this case, the $I$-band
magnitudes are accurate to 0.05 mag.

Completeness of our search for $\delta$~Sct variables depends on brightness
and amplitude of the stars. Precise long-term observations conducted from the
ground allow for a very effective detection of high-amplitude $\delta$~Sct
stars (HADS, $V$-band amplitudes $>0.15$ mag or $I$-band amplitudes $>0.1$ mag).
As we show in Section~4, our search is highly complete for variables
brighter than 17.5 mag and with amplitudes higher than 0.1 mag in the $I$ band.
Completeness of the search for bright HADS is similar to the completeness
for RRab type stars, that is of about 97\%.

%%%%%%%%%%%%%%%%%%%%%%%%%%%%%%%%%%%%%%%%%%%%%%%%%%%%%%%%%%%%%%%%%%%%

\Section{The Collection}

The OGLE collection of $\delta$~Sct-type variable stars in the
Galactic bulge fields contains 10~111 objects. Tables with basic parameters,
time-series $I$- and $V$-band photometry, and finding charts are available
to the astronomical community through the OGLE On-line Data Archive:
\begin{center}
{\it http://ogle.astrouw.edu.pl\\}
\end{center}
and
\begin{center}
{\it ftp://ftp.astrouw.edu.pl/ogle/ogle4/OCVS/blg/dsct/\\}
\end{center}
The stars are arranged according to increasing right ascension and named
as OGLE-BLG-DSCT-NNNNN, where NNNNN is a five-digit consecutive number.
In the data tables, we provide coordinates of the variables, the dominant
pulsation period, period uncertainty, and information on brightness.

Among the 10~111 detected $\delta$~Sct variables only two stars have a
GCVS designation (Samus \etal 2017). Variables OGLE-BLG-DSCT-06456 = V1363 Sgr
and OGLE-BLG-DSCT-06718 = V4117 Sgr were discovered by Blanco (1984)
and Gaposchkin (1955), respectively. The latter object is located in the
field of globular cluster NGC 6522 (variable V6, Clement \etal 2001)
in Baade's Window area, but the star probably does not belong to the cluster.
Other 42 $\delta$~Sct stars were identified in OGLE-I data
(Udalski \etal 1994, 1995ab, 1996, 1997). Seventy-six of the variables
were found by the MACHO team (Alcock \etal 2000). Another 139 sources
were later recognized in OGLE-II data by Pigulski \etal (2006). Seven $\delta$~Sct
pulsators were recently detected by the ASAS-SN survey (Jayasinghe \etal 2020).
Ten additional objects were found in the field of globular clusters M22
(7 stars, Kaluzny and Thompson 2001, Rozyczka \etal 2017) and M54 (3 stars,
Sollima \etal 2010), and classified as SX Phe stars. In total, 276 variables from
our collection were known before, or 9835 OGLE-IV $\delta$~Sct variables are
new discoveries. Other designations are also provided in the on-line data.

%%%%%%%%%%%%%%%%%%%%%%%%%%%%%%%%%%%%%%%%%%%%%%%%%%%%%%%%%%%%%%%%%%%%%

\Section{General Properties of the Detected $\delta$~Sct Stars}

The new collection is a source of various information on the $\delta$~Sct
stars themselves and stellar populations they belong to. In this section,
we present some general properties of the detected stars.

Fig.~1 shows the distribution of the $\delta$~Sct variables overlaid onto
the contours of the 121 investigated OGLE-IV Galactic bulge fields.
Due to the increasing interstellar extinction toward the Galactic plane,
the variables get fainter and they are practically not observed at low
latitudes ($|b|<1\arcd$) in the optical range. The stars concentrate
toward the Galactic center. This suggests that most of the detected
$\delta$~Sct objects are located at distances of several kpc or more from
us and that they mainly belong to the intermediate-age and old populations.
Some of the stars are likely metal-poor representatives of the old disk
and halo components and could be classified as SX~Phe-type variables.
Nevertheless, most of the variables seem to belong to the Galactic bulge.

\begin{figure}[htb!]
\centerline{\includegraphics[angle=0,width=130mm]{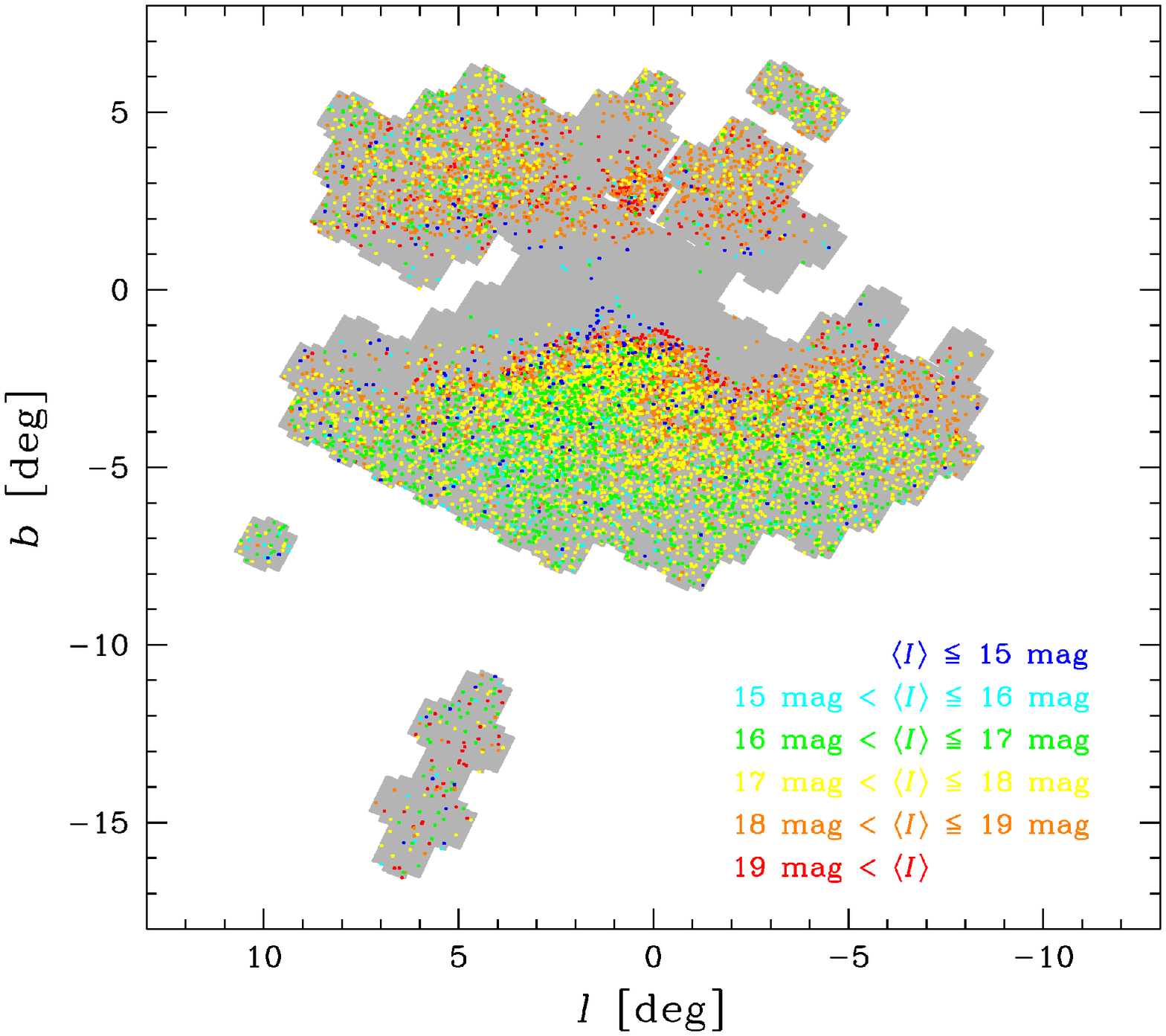}}
\FigCap{Distribution, in Galactic coordinates, of 10~111 $\delta$~Sct-type
variable stars detected in 121 OGLE-IV bulge fields (marked in grey) spreading
over 172 deg$^2$. Seven fields located around $(l,b)=(+5\arcd,-13\arcd)$
cover the central part of the Sgr dSph galaxy. A single outlying field at
$(l,b)\approx(+10\arcd,-7\arcd)$ includes Galactic globular cluster M22.
The colors code mean $I$-band brightness of the variables.}
\end{figure}

For variables with available $V$-band data it was possible to construct
an observed color-magnitude diagram (see Fig.~2). The majority of $\delta$~Sct
stars are smeared in a lane parallel to the reddening vector (determined from the
OGLE bulge RR Lyr stars, Pietrukowicz \etal 2015). Stars above the lane (brighter)
are foreground, likely halo and thick disk objects. Stars below the lane (fainter)
forming a vertical sequence at $V-I\approx0.5$ mag, belong to the Sgr dSph galaxy
which core, the globular cluster M54, is located at the distance of 26.7 kpc
(\eg Hamanowicz \etal 2016). It is worth noting that the on-sky distribution
and the color-magnitude diagram for $\delta$~Sct stars are very similar to
those of bulge RR Lyr variables (see Fig.~2 in Soszy\'nski \etal 2014
and Fig.~1 in Pietrukowicz \etal 2015, respectively).

\begin{figure}[htb!]
\centerline{\includegraphics[angle=0,width=110mm]{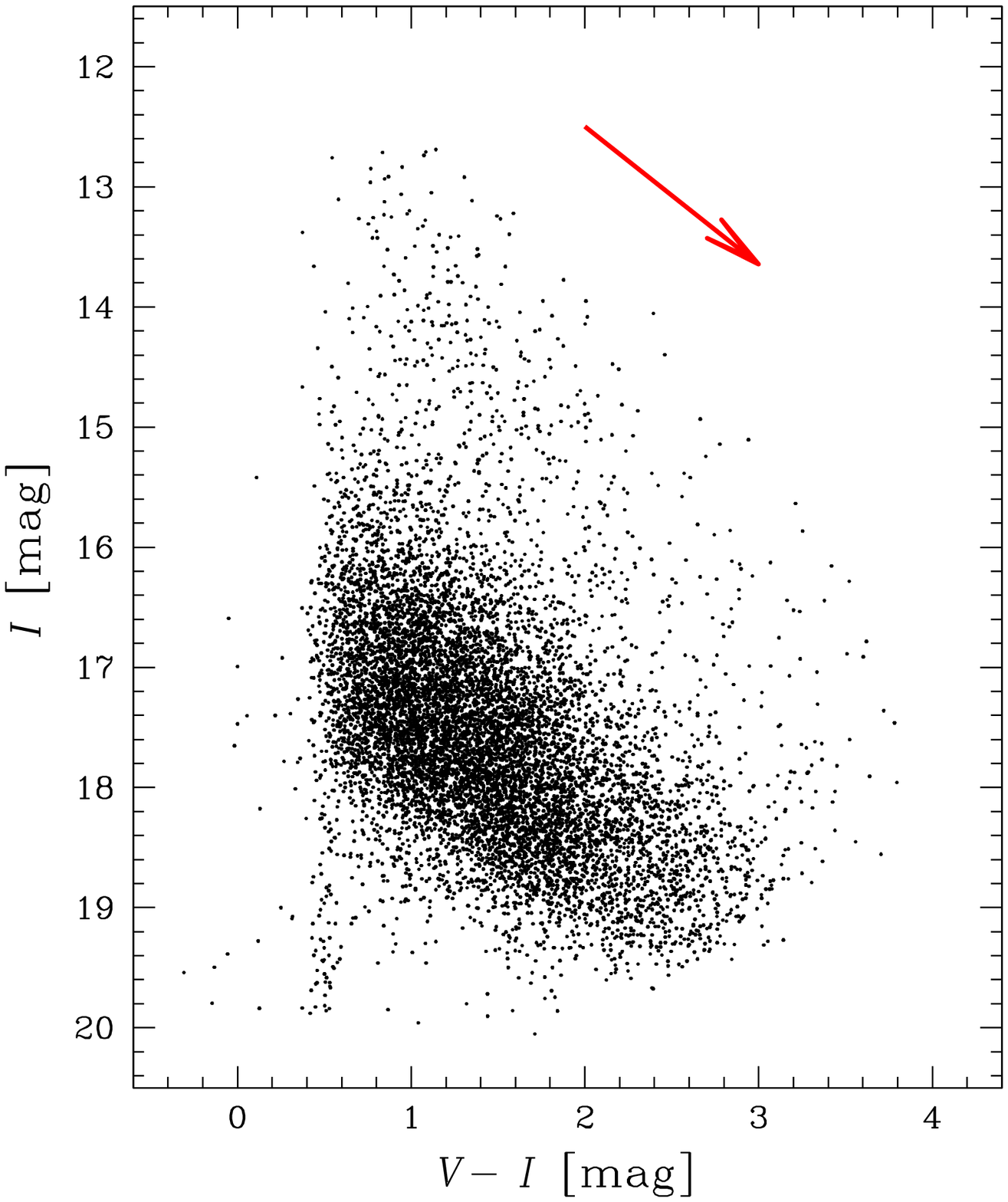}}
\FigCap{($V-I$) \vs $I$ diagram for 9579 $\delta$~Sct stars. The arrow
represents the reddening vector in the bulge area. Points forming a
sequence around the color $V-I=0.5$~mag in the brightness range
$18<I<20$~mag are background $\delta$~Sct stars from the Sgr dSph galaxy.}
\end{figure}

Fig.~3 contains three panels with, from top to bottom, the mean
$I$-band brightness, pulsation period, and peak-to-peak $I$-band amplitude
distributions of detected $\delta$~Sct stars. The amplitudes were
determined for 2880 single-mode pulsators only. Based on the brightness
and amplitude distributions we can infer that the presented collection is
highly complete for variables with mean $I$-band brightness $<17.5$~mag
and with amplitudes higher than 0.1 mag. There is only one maximum
in the period distribution, at about 0.07~d for the brighter variables
($I\leq17.5$~mag).\footnote{In the whole collection, there is only one
variable with a pulsation period below 0.04~d (OGLE-BLG-DSCT-09299 with
$P_{\rm puls}=0.0399$~d) and thus it is not seen in the logarithmic scale
histogram in Fig.~3. Many inspected variable sources with periods below
0.04~d and not classified by us as BLAPs have sinusoidal or nearly
sinusoidal light curve shapes. Those objects could be SX Phe variables
as well as pulsating variables of other types (\eg subdwarfs, white dwarfs).
We plan to analyze the shortest-period objects separately.}
It is impossible to separate shorter-period metal-poor Population II (SX~Phe)
stars from longer-period metal-rich Population I $\delta$~Sct stars
by the pulsation period. The former pulsators are likely much less abundant
in the whole sample.

Period--amplitude diagram is shown in Fig.~4. There are no characteristic
features in this diagram, in contrast to such (Bailey) diagram for RR Lyr
stars (Soszy\'nski \etal 2014).

\begin{figure}[htb!]
\centerline{\includegraphics[angle=0,width=120mm]{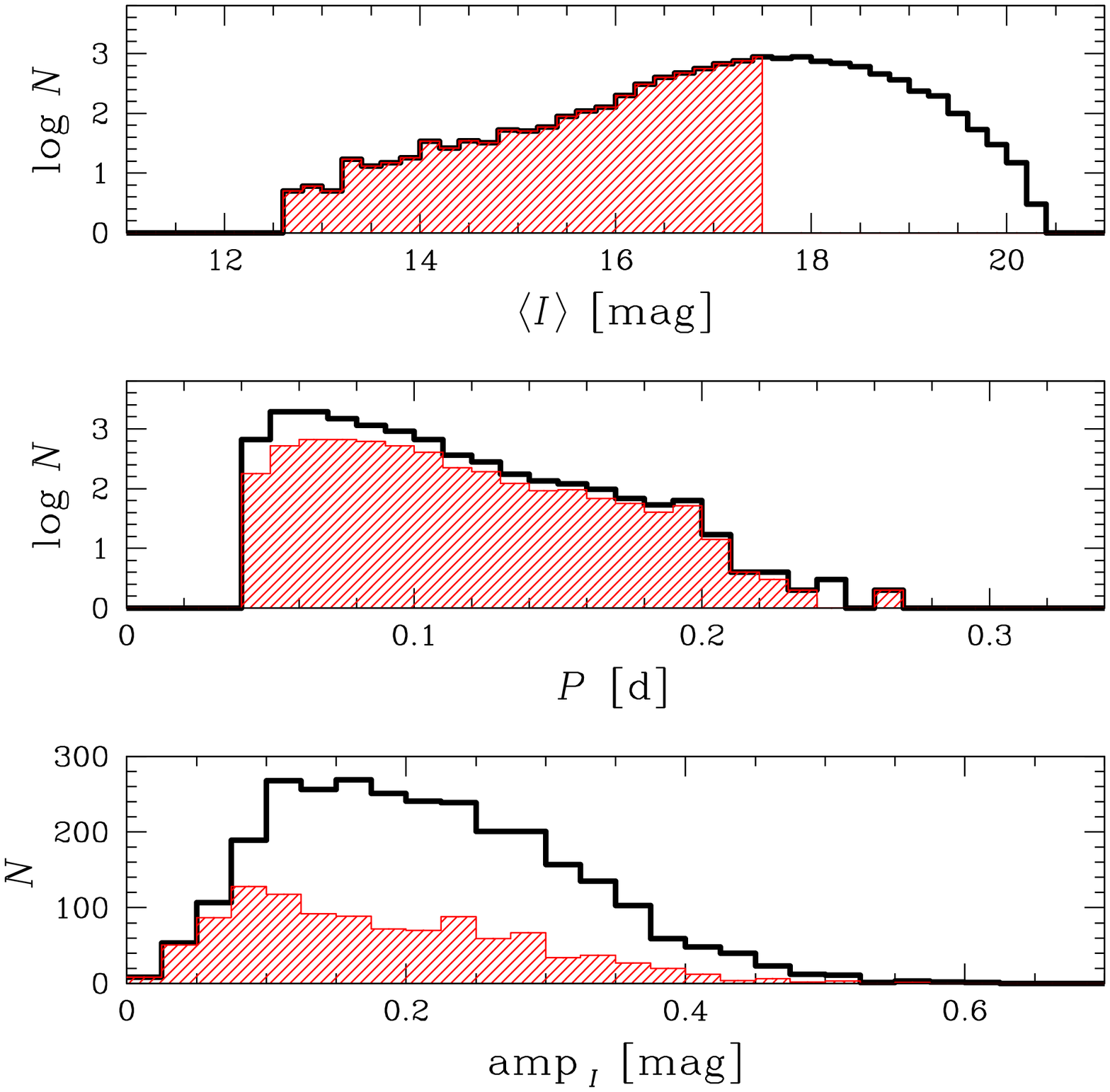}}
\FigCap{{\it Upper panel}: Mean $I$-band brightness distribution for the
whole set of 10~111 OGLE-IV $\delta$~Sct variables. The bin size is 0.2~mag.
The vertical axis has a logarithmic scale. High completeness of the sample
reaches a magnitude of $I=17.5$ (shaded in red). {\it Middle panel}: Histogram
of the dominant period for all detected 10~111 $\delta$~Sct stars (black line)
and 4533 variables brighter than $\langle I\rangle=17.5$ mag (in red).
The bin size is 0.01~d. The vertical axis has a logarithmic scale.
{\it Lower panel}: Histogram of the peak-to-peak $I$-band amplitude
determined for all 2880 single-mode $\delta$~Sct pulsators (black line)
and 1075 single-mode pulsators with $\langle I\rangle<17.5$ mag (shaded
in red). The bin size is 0.025~mag.}
\end{figure}

\begin{figure}[htb!]
\centerline{\includegraphics[angle=0,width=110mm]{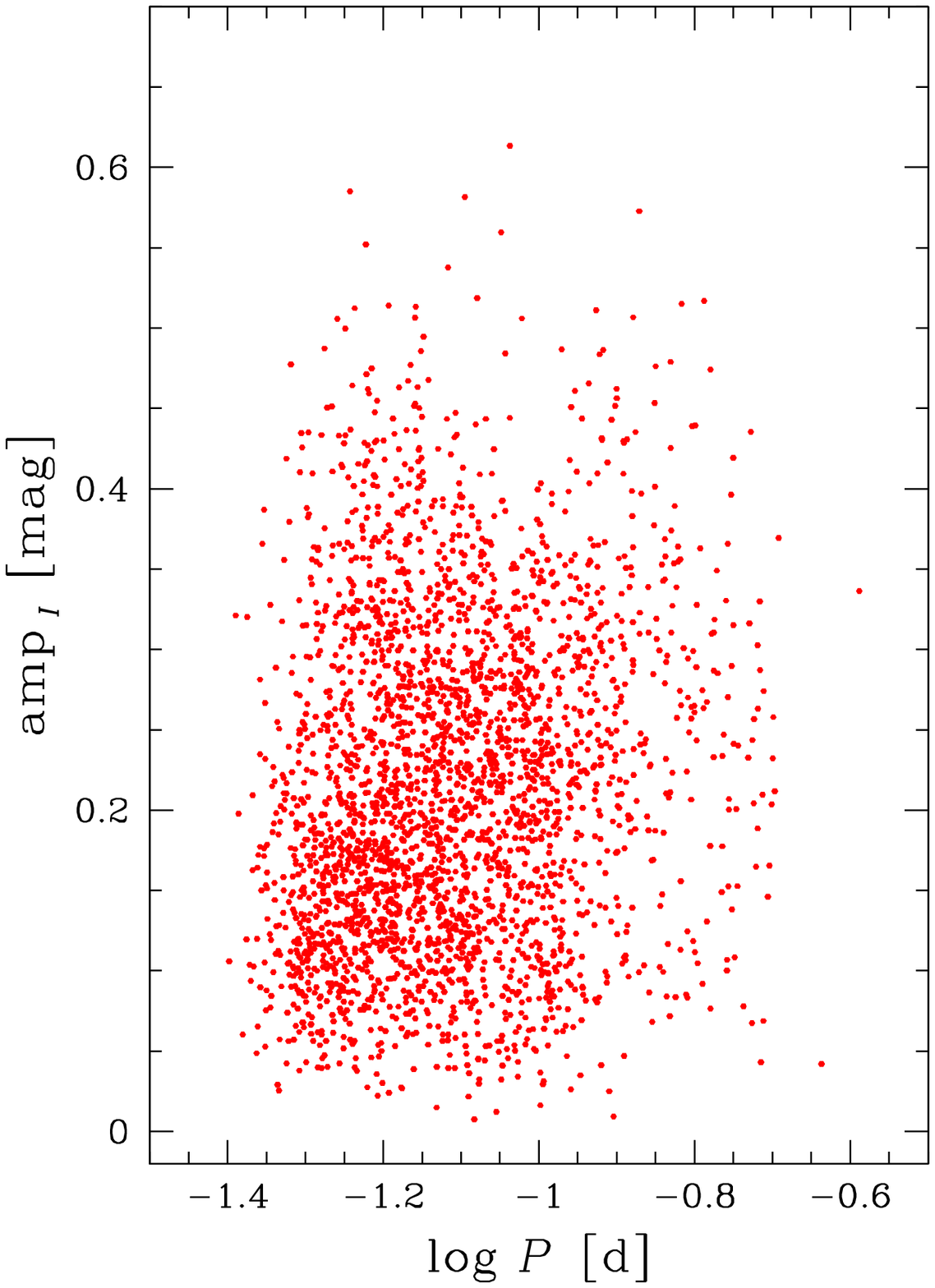}}
\FigCap{Period--amplitude diagram for 2880 single-mode $\delta$~Sct stars.
There are no evident features in this diagram.}
\end{figure}

We decomposed the light curves of single-mode pulsators into cosine
Fourier series. In Fig.~5, we plot the Fourier coefficient combinations,
amplitude ratio $R_{21}$ and $R_{31}$, and phase combinations
$\phi_{21}$ and $\phi_{31}$, in the function of period. The phase
combinations clearly correlate with the period.

\begin{figure}[htb!]
\centerline{\includegraphics[angle=0,width=130mm]{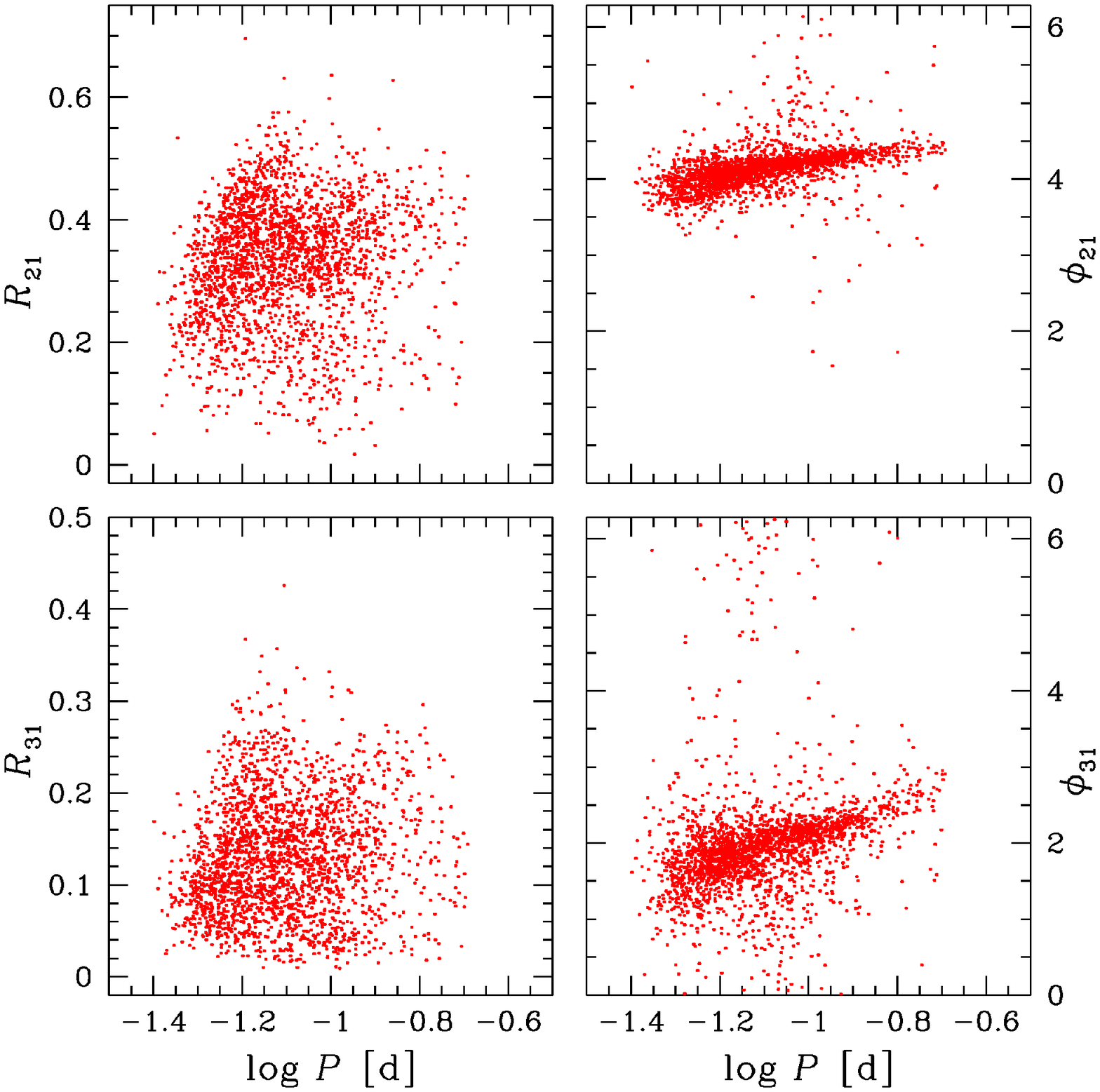}}
\FigCap{Fourier coefficient combinations $R_{21}$, $R_{31}$, $\phi_{21}$,
and $\phi_{31}$ as a function of the logarithm of the pulsation period
for single-mode $\delta$~Sct stars. Evident correlations are seen
for both Fourier phase combinations.}
\end{figure}

%%%%%%%%%%%%%%%%%%%%%%%%%%%%%%%%%%%%%%%%%%%%%%%%%%%%%%%%%%%%%%%%%%%%%

\Section{Possible Members of Globular Clusters}

Some of the variables in this collection belong or likely belong to
globular clusters. We scanned the whole sample for $\delta$~Sct/SX Phe
stars in the area of 37 globular clusters observed in the OGLE-IV bulge fields.
In Table~1, we present a list of 22 variables located within three half-light
radii ($r<3r_{\rm h}$) from the centers of nine clusters. We used
a brightness criterion that SX Phe stars are about 1.7--3.3 mag fainter
in $V$ than the horizontal branch (HB) of the cluster (\eg Kaluzny \etal 1996,
Kaluzny and Thompson 2001). Parameters of the clusters (coordinates,
radii, HB magnitudes) were taken from Harris (1996) catalog (version 2010).
Seven objects in globular cluster M22 (NGC 6656) were previously known
(Kaluzny and Thompson 2001, Rozyczka \etal 2017) and are
confirmed members (Zloczewski \etal 2012, Rozyczka \etal 2017).
The remaining 15 variables could belong to the clusters. We performed a similar
scanning operation for the radius $r<5r_{\rm h}$ finding 35 objects.
However, one has to remember that most of the observed clusters are
immersed in relatively dense bulge environment and separating
bulge from cluster stars can be problematic at larger distances from
the cluster centers.

\begin{table}[htb!]
\centering \caption{\small OGLE-IV $\delta$~Sct/SX Phe variables located
within the radius of $3r_{\rm h}$ from the centers of globular clusters}
\medskip
{\small
\begin{tabular}{lcccccl}
\hline
Cluster   & Variable & $\langle I\rangle$ & $\langle V\rangle$ & $P_{\rm puls}$ & $r/r_{\rm h}$ & Other name\\
          & OGLE-BLG- &        &         &          &               & \\
          & -DSCT-   &  [mag]  &  [mag]  &    [d]   &               & \\
\hline
NGC 6304  &   00038  &  17.00  &  17.95  &  0.08804739(1)  &  2.30  & \\
          &   00055  &  17.63  &  18.64  &  0.08810378(2)  &  1.71  & \\
          &   00057  &  17.77  &  18.78  &  0.06559091(1)  &  2.45  & \\
NGC 6401  &   01179  &  18.57  &  20.28  &  0.05328611(1)  &  2.80  & \\
          &   01209  &  18.22  &  19.73  &  0.10545024(4)  &  1.26  & \\
          &   01243  &  18.77  &  20.18  &  0.06473004(1)  &  2.46  & \\
Pal 6     &   01905  &  18.68  &  21.37  &  0.10669881(4)  &  1.86  & \\
NGC 6453  &   03154  &  18.57  &  19.72  &  0.05319618(1)  &  2.03  & \\
Djorg 2   &   06270  &  18.27  &  19.88  &  0.04707594(1)  &  2.50  & \\
NGC 6544  &   07604  &  15.68  &  17.10  &  0.08068763(1)  &  0.48  & \\
          &   07630  &  16.03  &  17.46  &  0.04352380(1)  &  0.95  & \\
NGC 6558  &   08161  &  17.35  &  18.25  &  0.05704527(1)  &  2.47  & \\
          &   08283  &  17.13  &  18.02  &  0.05511362(2)  &  2.36  & \\
NGC 6569  &   08825  &  18.58  &  19.51  &  0.05483726(4)  &  1.00  & \\
M22       &   09961  &  15.98  &  16.81  &  0.04731774(4)  &  0.59  & KT-34$^1$ \\
          &   09963  &  15.66  &  16.42  &  0.04432495(2)  &  0.50  & KT-29$^1$ \\
          &   09964  &  15.31  &  16.17  &  0.05560342(7)  &  0.66  & KT-28$^1$ \\
          &   09965  &  15.81  &  16.63  &  0.05007784(2)  &  0.18  & KT-45$^1$ \\
          &   09966  &  16.08  &  16.89  &  0.04217440(3)  &  0.29  & KT-27$^1$ \\
          &   09967  &  15.54  &  16.33  &  0.08364628(3)  &  1.03  & KT-54$^1$ \\
          &   09968  &  14.86  &  15.94  &  0.06231611(2)  &  0.12  & V112$^2$  \\
          &   09969  &  15.68  &  16.44  &  0.1077372(2)   &  1.39  & \\
\hline
\end{tabular}}\\
\smallskip
{\footnotesize Mean brightness, pulsation period, and distance from the cluster center
are provided\\ for each variable. Seven members of globular cluster M22 were
previously reported in\\ $^1$Kaluzny and Thompson (2001), and $^2$Rozyczka \etal (2017)}\\
\end{table}

%%%%%%%%%%%%%%%%%%%%%%%%%%%%%%%%%%%%%%%%%%%%%%%%%%%%%%%%%%%%%%%%%%%%%

\Section{Peculiarities in the $\delta$~Sct Light Curves}

In this section, we illustrate the variety of light curve shapes
of the $\delta$~Sct pulsators. Some of the presented objects
resemble variables of other types. Very often, the shape, period,
and peak-to-peak amplitude allow for unambiguous classification.
In some cases, however, to confirm that the variables are indeed of
the $\delta$~Sct type, we additionally verified their positions
in color-magnitude diagrams constructed for stars from the same field.
The $\delta$~Sct stars reside in the area around the upper or middle
MS in the observed diagrams.

In Fig.~6, we show four examples of high-amplitude $\delta$~Sct
stars with nearly flat minima. The minima cover up to about 0.4
of the cycle. High amplitudes and sharp maxima point to the
fundamental mode. Variables with such light curve
shape have periods in the range roughly 0.05--0.07 d.

Fig.~7 presents four $\delta$~Sct variables with sharp maxima and
almost symmetric shape. Such shape resembles light curves
of binary systems with a prominent reflection effect due to large
difference in effective temperature between the components (for instance,
in hot subdwarf--brown dwarf binaries). The mild asymmetry at the period
of a few hours clearly indicates that these objects are pulsating
$\delta$~Sct stars. The round minima and low amplitudes of
$\approx0.1$~mag point to an overtone mode.

\begin{figure}[htb!]
\centerline{\includegraphics[angle=0,width=130mm]{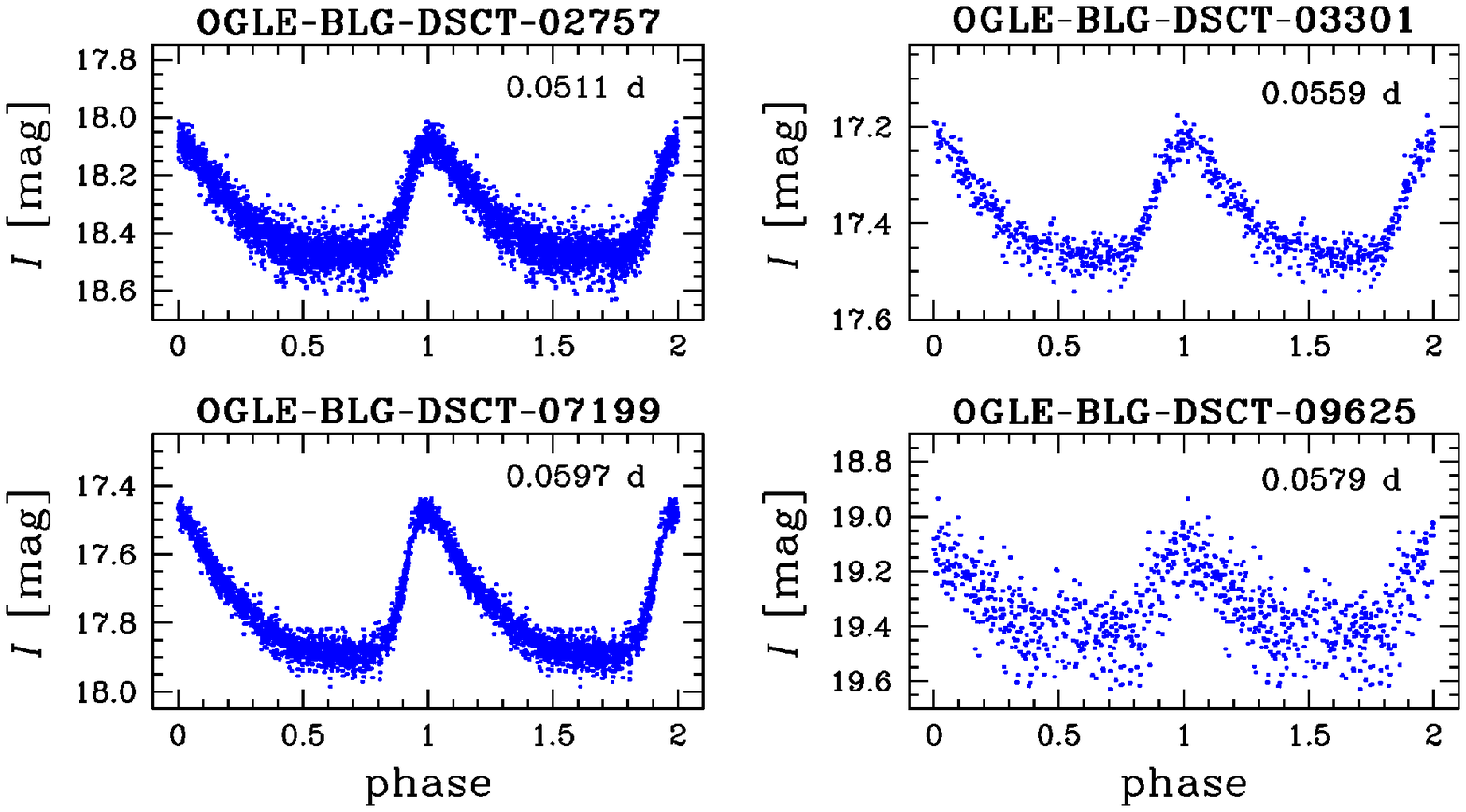}}
\FigCap{Phased $I$-band light curves of selected $\delta$~Sct variables
with wide, nearly flat minima.}
\end{figure}

\begin{figure}[htb!]
\centerline{\includegraphics[angle=0,width=130mm]{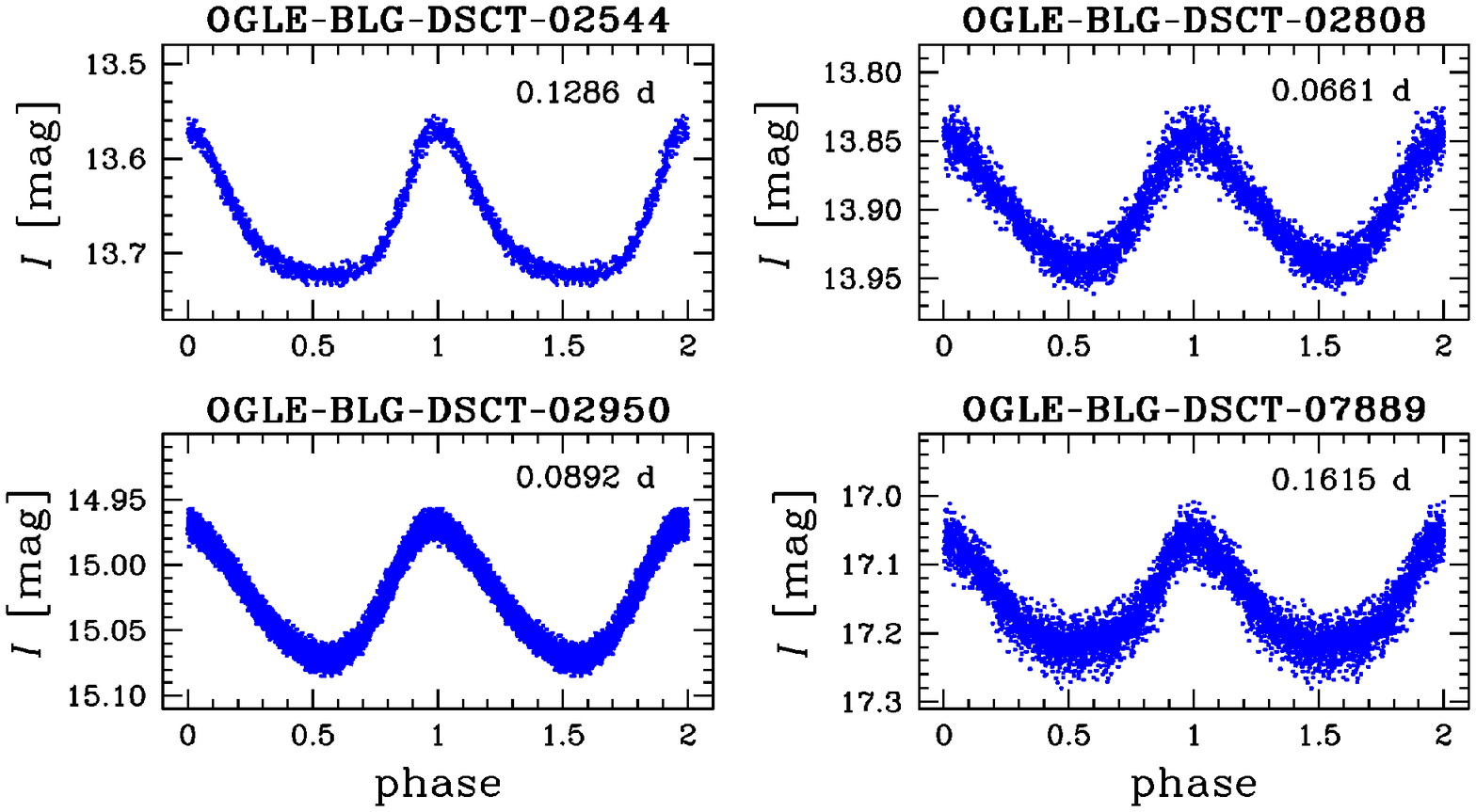}}
\FigCap{Examples of $\delta$~Sct variables with nearly symmetric light curves.}
\end{figure}

In Fig.~8, we present another group of nearly symmetric light curves
of $\delta$~Sct variables. This group mimics contact binary systems
if double the period. Nevertheless, even the doubled period is shorter
than the orbital period observed in almost all contact binaries ($>0.2$~d).
$I$-band amplitudes of such $\delta$~Sct variables are of about 0.1~mag
or smaller.

\begin{figure}[htb!]
\centerline{\includegraphics[angle=0,width=130mm]{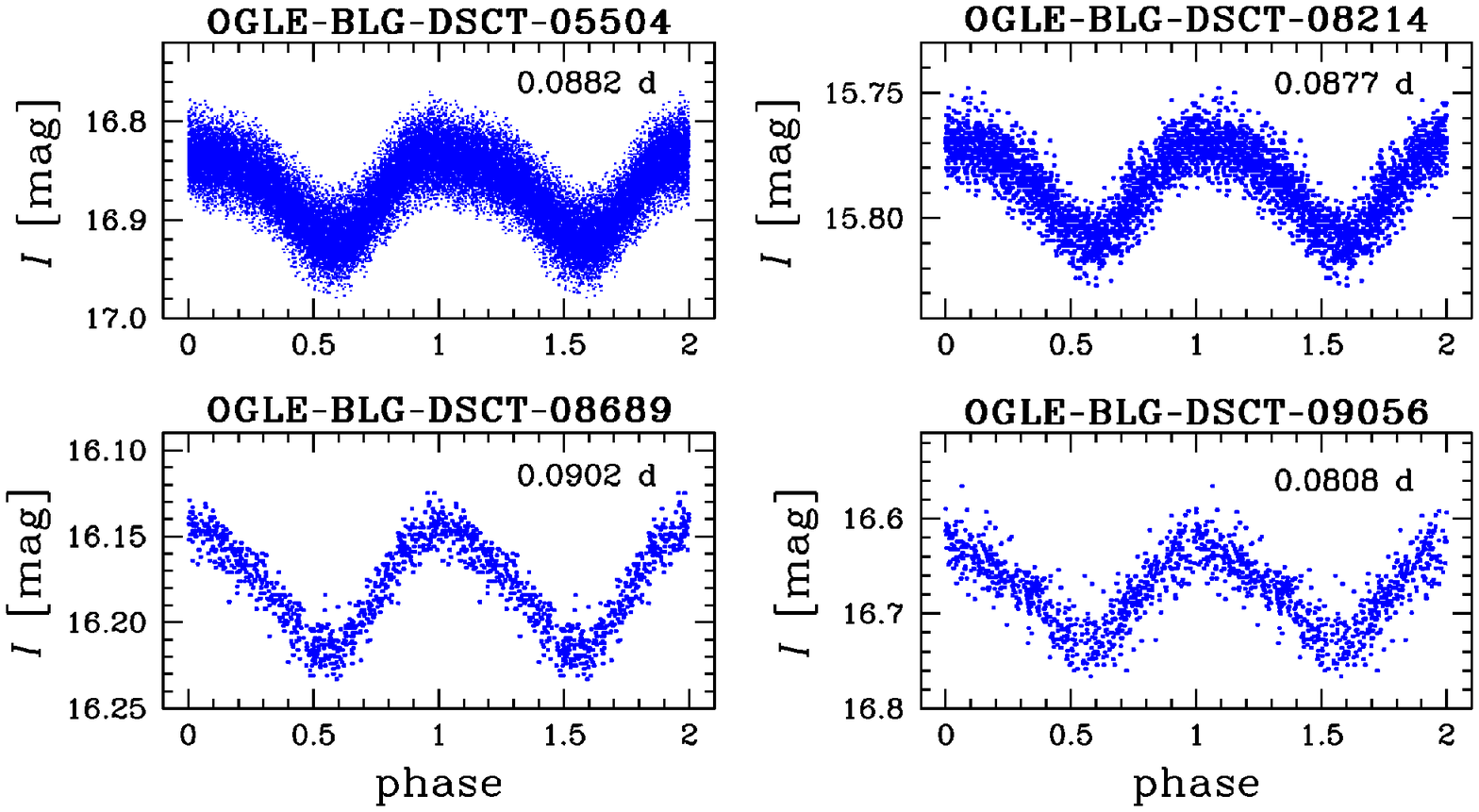}}
\FigCap{Examples of $\delta$~Sct variables that mimic contact binary
stars when double the given period.}
\end{figure}

In the following figure, Fig.~9, one can find $\delta$~Sct variables
showing reverse brightness variation to typical pulsating
stars---the time from the minimum to the maximum is longer
than the time from the maximum to the minimum. An identical shape, 
with a broken rising branch half-way to the maximum light, was
already found in triple-mode classical Cepheid OGLE-SMC-CEP-1350
(see Fig.~5 in Soszy\'nski \etal 2010). Light curve decomposition
of that star pointed to the first overtone. By analogy, we suspect
that the bulge objects presented in Fig.~9 are first-overtone pulsators.

\begin{figure}[htb!]
\centerline{\includegraphics[angle=0,width=130mm]{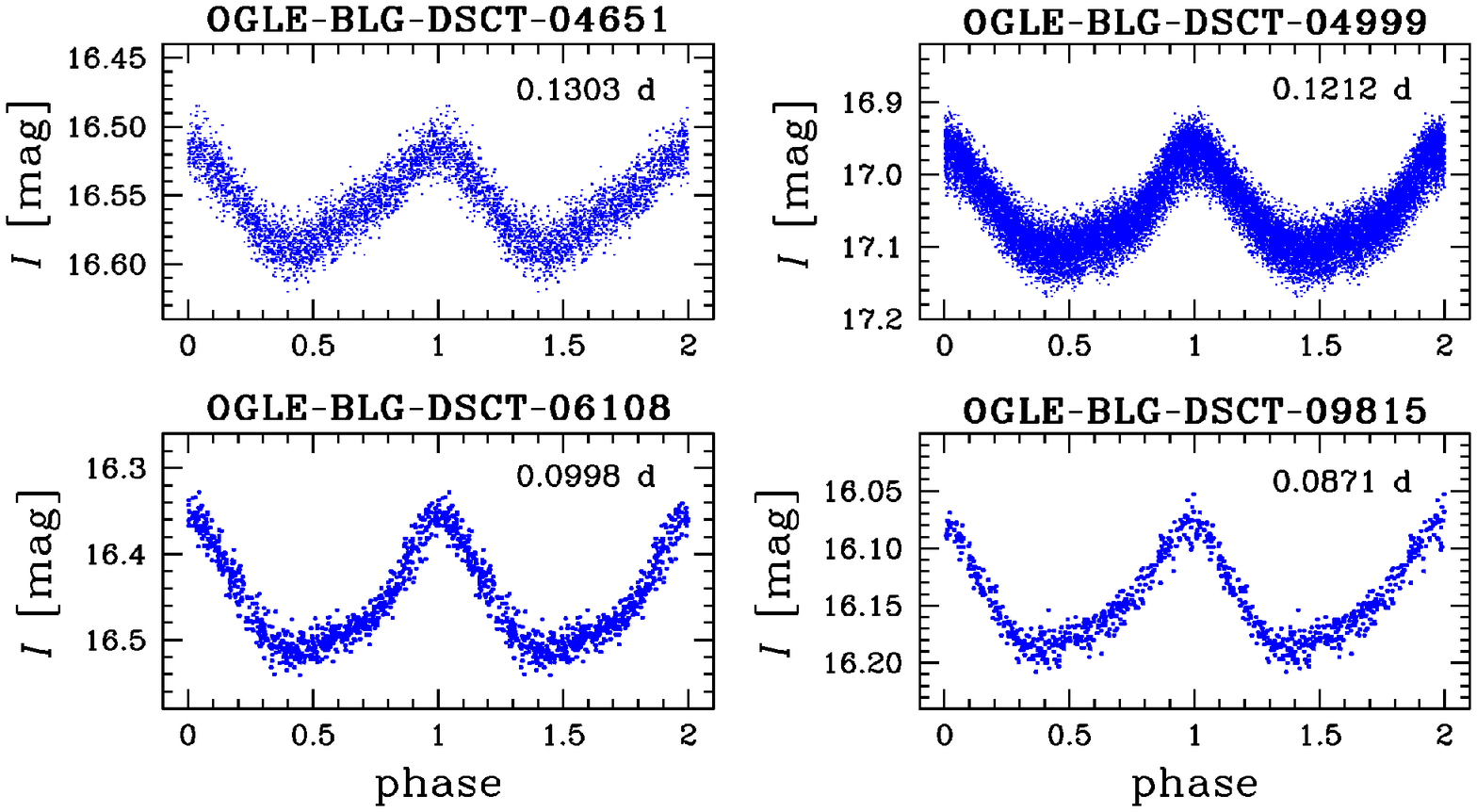}}
\FigCap{Examples of phased light curves of $\delta$~Sct variables with
the rising part lasting longer than the fading one. These are likely
first overtone pulsators.}
\end{figure}

In Fig.~10, we present examples of $\delta$~Sct light curves with
sharp extrema. Fig.~11 shows variables with pronounced additional
bumps on both the rising and fading branches. Such light curve shapes
are observed in the period range from about 0.07 to 0.09~d.

\begin{figure}[htb!]
\centerline{\includegraphics[angle=0,width=130mm]{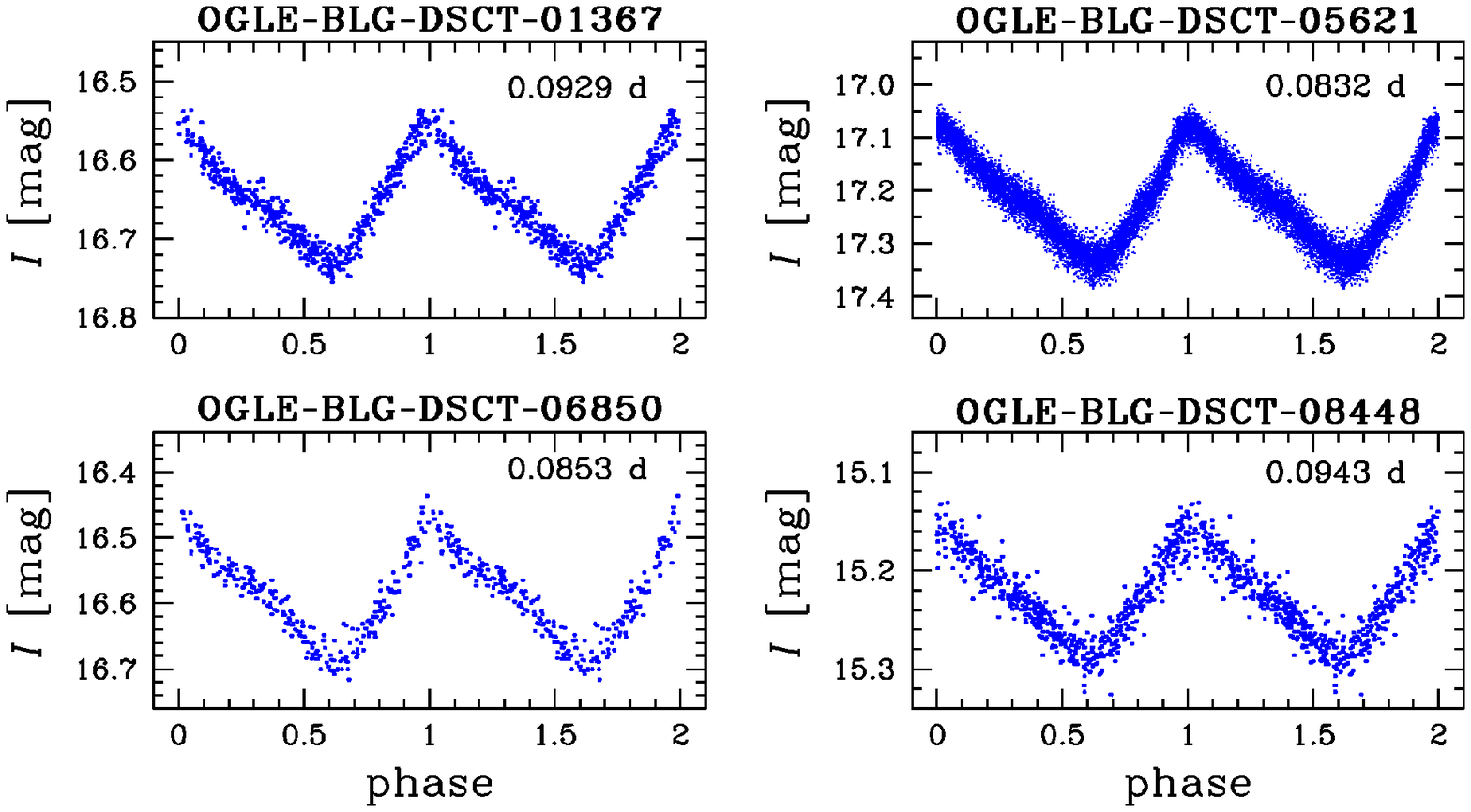}}
\FigCap{Light curves of $\delta$~Sct variables with sharp extrema.}
\end{figure}

\begin{figure}[htb!]
\centerline{\includegraphics[angle=0,width=130mm]{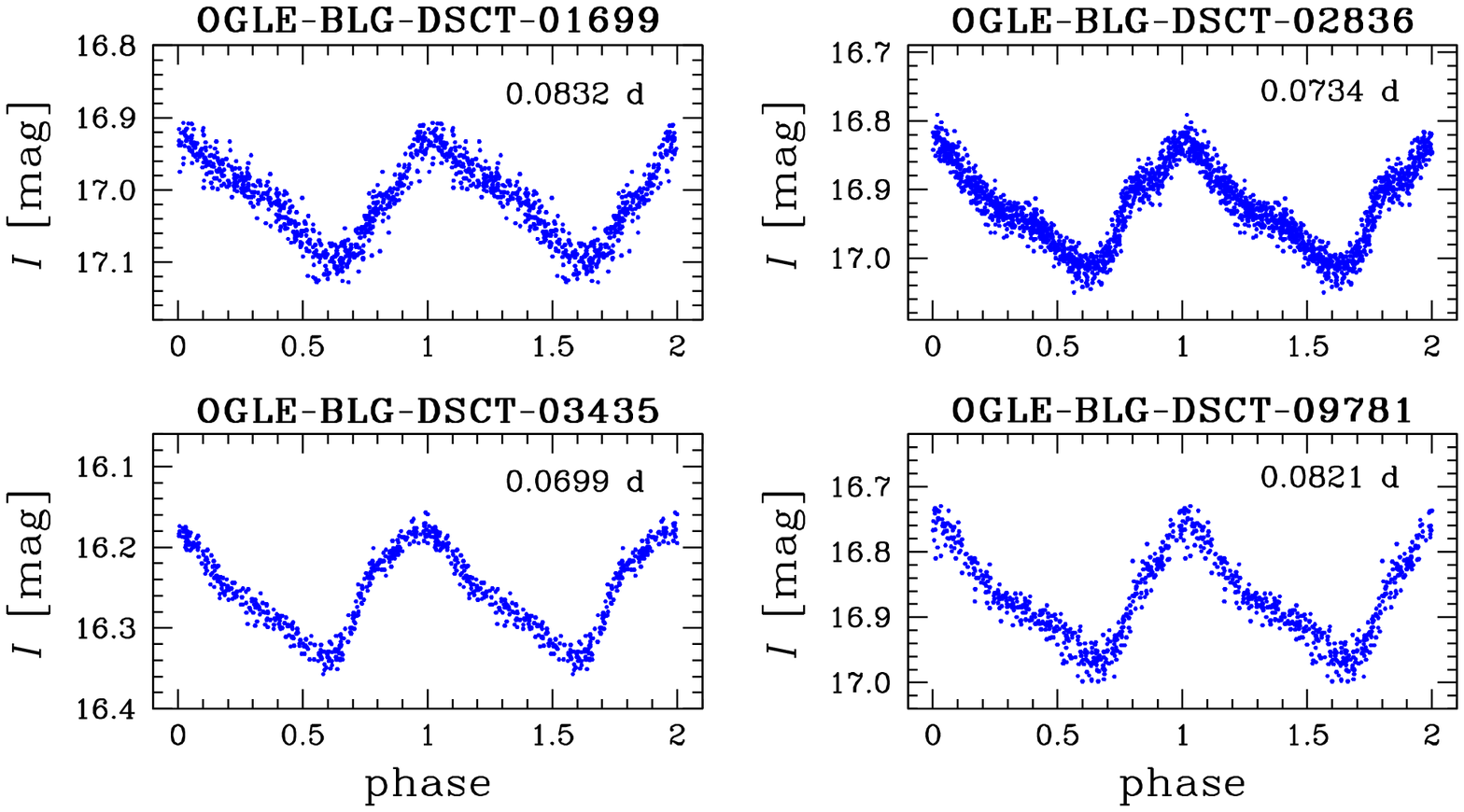}}
\FigCap{$\delta$~Sct variables with bumps on the rising and fading branches.}
\end{figure}

In Fig.~12, we compare phased, saw-tooth-shaped light curves
of presumably fundamental-mode $\delta$~Sct pulsators in a wide period
range of 0.05--0.2~d. Amplitudes of the stars were not scaled.
One can follow the evolution of the light curve shape with the
increasing period similar to the Hertzsprung progression in Cepheids.
The light curve of a 0.05-d fundamental-mode $\delta$~Sct
star is very similar in shape to a 1.0-d classical Cepheid and 0.9-d
anomalous Cepheid showing a small bump around the minimum light.
The shape of a 0.2-d $\delta$~Sct star resembles the shape of a 5-d
classical Cepheids with the bump appearing on the fading branch.

\begin{figure}[htb!]
\centerline{\includegraphics[angle=0,width=130mm]{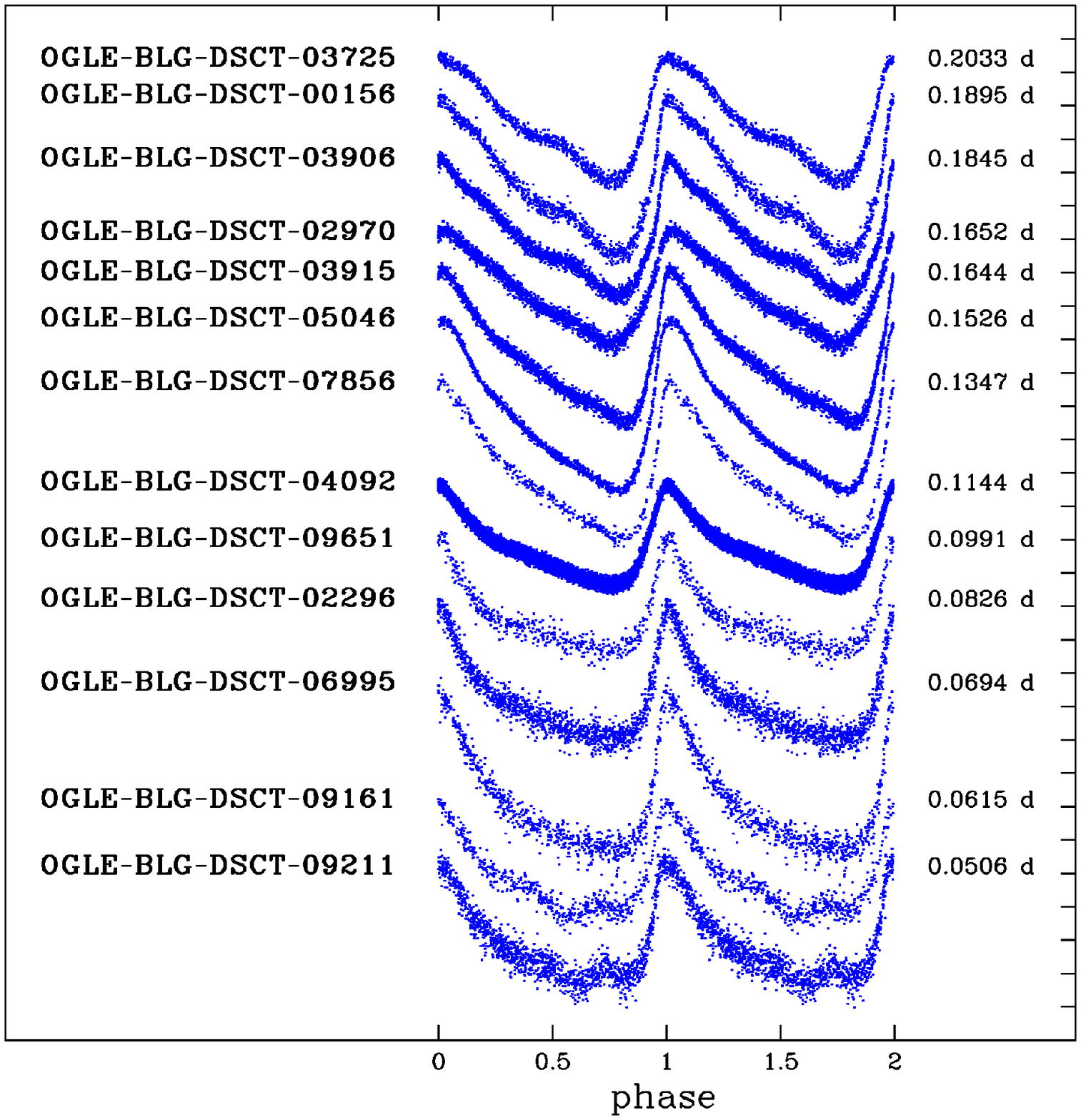}}
\FigCap{Light curve shape changes with the pulsation period in
fundamental-mode $\delta$~Sct stars. The light curves are shifted to
avoid overlaps, but the amplitudes are preserved. The ticks on the vertical
axis are every 0.1~mag.}
\end{figure}

During the visual inspection of $\delta$~Sct light curves, we encountered
many evident multi-mode pulsators. Detailed analysis of additional
periodicities is the topic of the work by Netzel \etal (2021, in prep.).
Here in Fig.~13, we show three double-mode $\delta$~Sct stars
pulsating in the fundamental mode (F) and first overtone (1O), simultaneously.
The mode types are confirmed by relatively high amplitudes of 0.1--0.2 mag
and the period ratio $P_{\rm 1O}/P_{\rm F}\approx0.77$ (see sequences
in the Petersen diagram in Fig.~10 in Pietrukowicz \etal 2013).
In the power spectrum, the presented stars also show a combination
frequency $f=f_{\rm 1O}-f_{\rm F}$, which proves the intrinsic origin
of the double-mode signal. We would like to pay attention to
the shape of the first-overtone component in star OGLE-BLG-DSCT-00875.
Light curves of a similar shape with round symmetric minima, are observed
in some other $\delta$~Sct stars (see Fig.~14). In these stars,
the dominant mode seems to be the first overtone.

\begin{figure}[htb!]
\centerline{\includegraphics[angle=0,width=130mm]{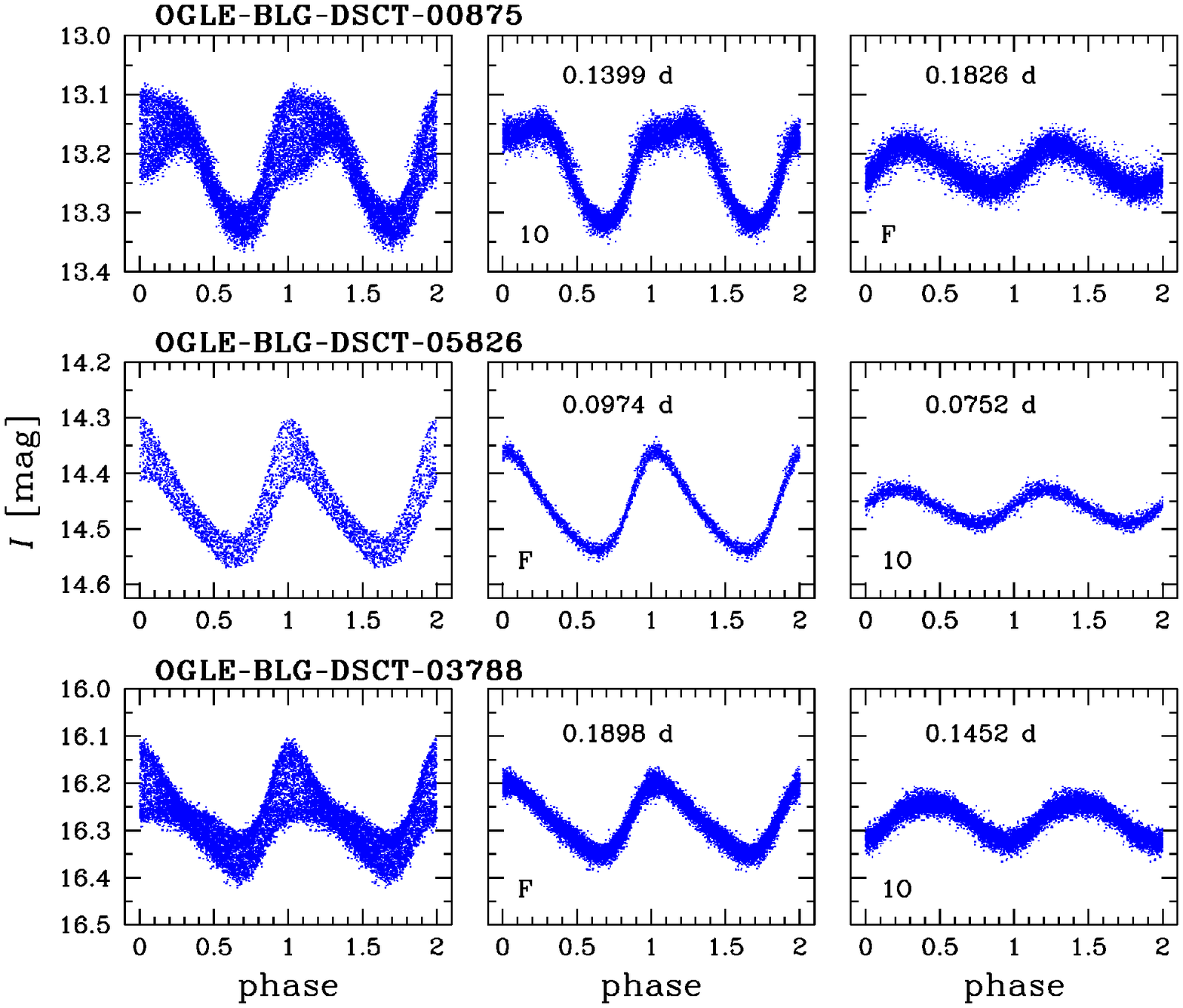}}
\FigCap{Three examples of double-mode (fundamental + first overtone)
$\delta$~Sct pulsators. Note an unusual shape of the first overtone
component in star OGLE-BLG-DSCT-00875 ({\it top middle panel}).}
\end{figure}

\begin{figure}[htb!]
\centerline{\includegraphics[angle=0,width=130mm]{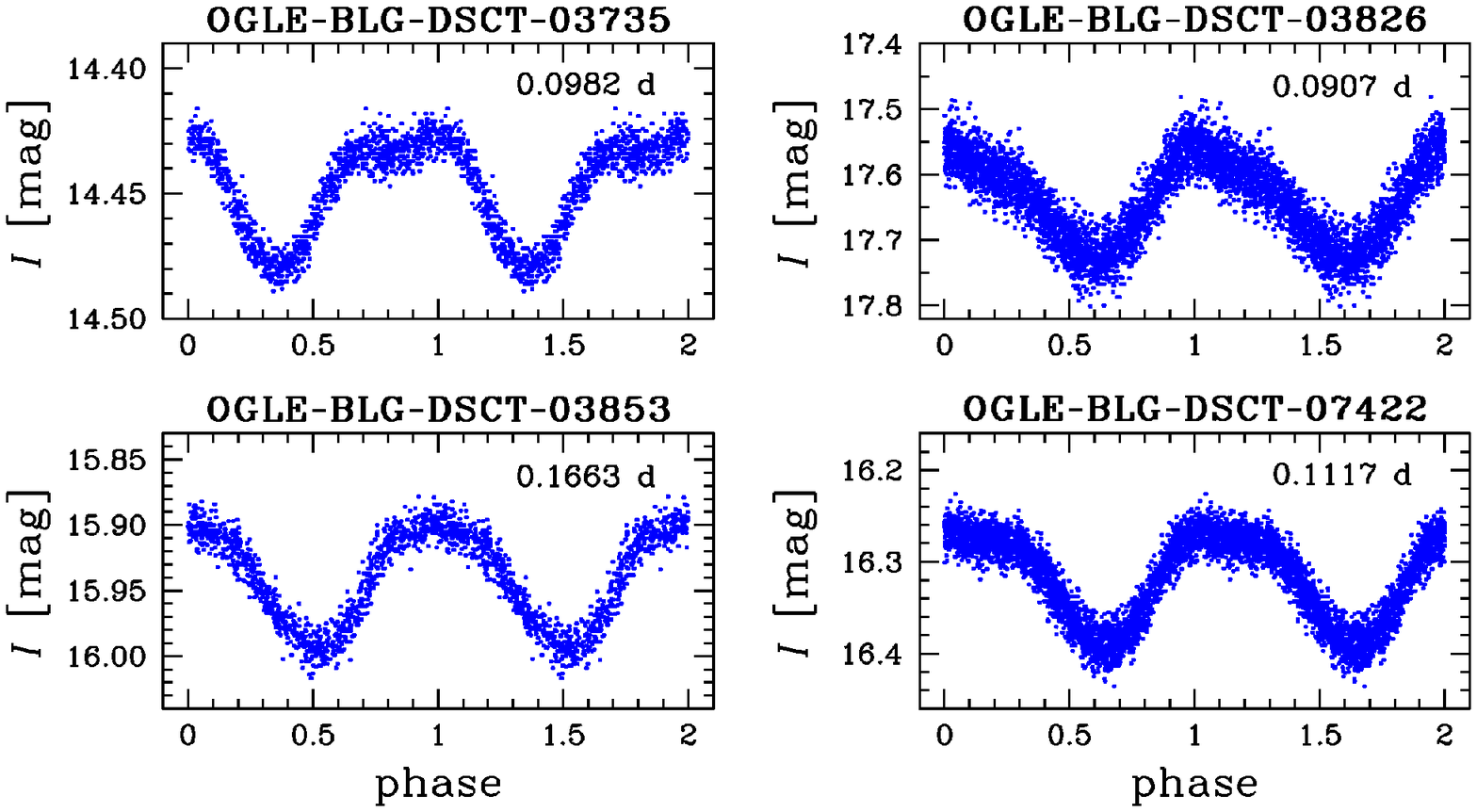}}
\FigCap{Four examples of first-overtone $\delta$~Sct pulsators showing
the unusual shape encountered in OGLE-BLG-DSCT-00875 (see Fig.~13).}
\end{figure}

In Fig.~15, we show a few examples of $\delta$~Sct stars with
mean brightness variations observed over the whole decade of OGLE-IV.
In object OGLE-BLG-DSCT-05430, for instance, the mean $I$-band
brightness dropped monotonically by 0.15~mag. In star OGLE-BLG-DSCT-07133,
the mean brightness varied irregularly over the years 2010--2019
reaching an amplitude of 0.053~mag. The observed variations may
stem from the presence of an active companion to the pulsating star.

\begin{figure}[htb!]
\centerline{\includegraphics[angle=0,width=130mm]{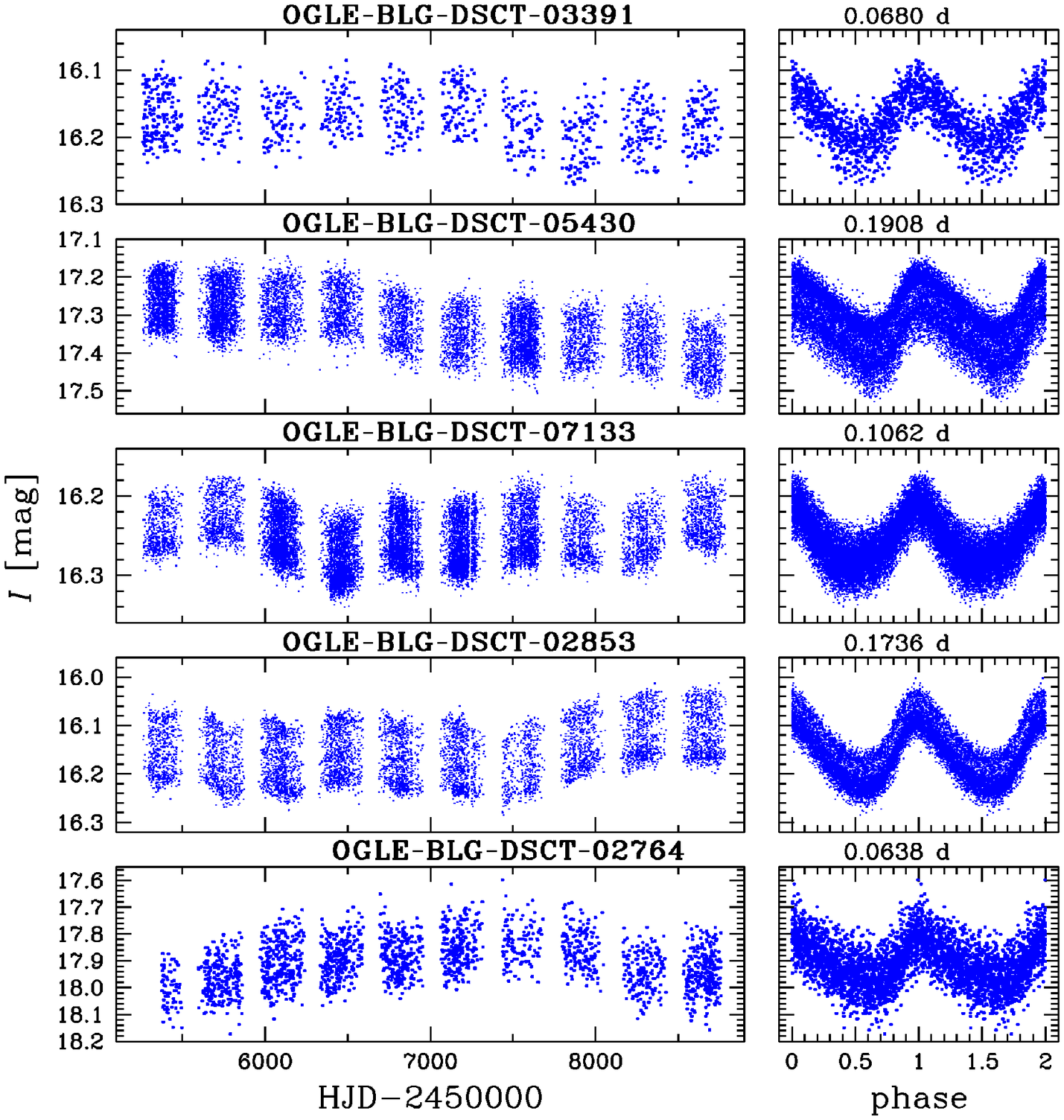}}
\FigCap{Examples of $\delta$~Sct variables showing mean brightness
variations over the years 2010--2019.}
\end{figure}

Some of the $\delta$~Sct stars exhibit noticeable period changes.
Fig.~16 contains four such examples. In each panel, we overplot
the phased light curve from 2014 onto the full OGLE-IV light curve.
In the case of object OGLE-BLG-DSCT-06232, the changes seem
to be large. A detailed analysis of period changes in $\delta$~Sct
stars is planned in a separate article.

\begin{figure}[htb!]
\centerline{\includegraphics[angle=0,width=130mm]{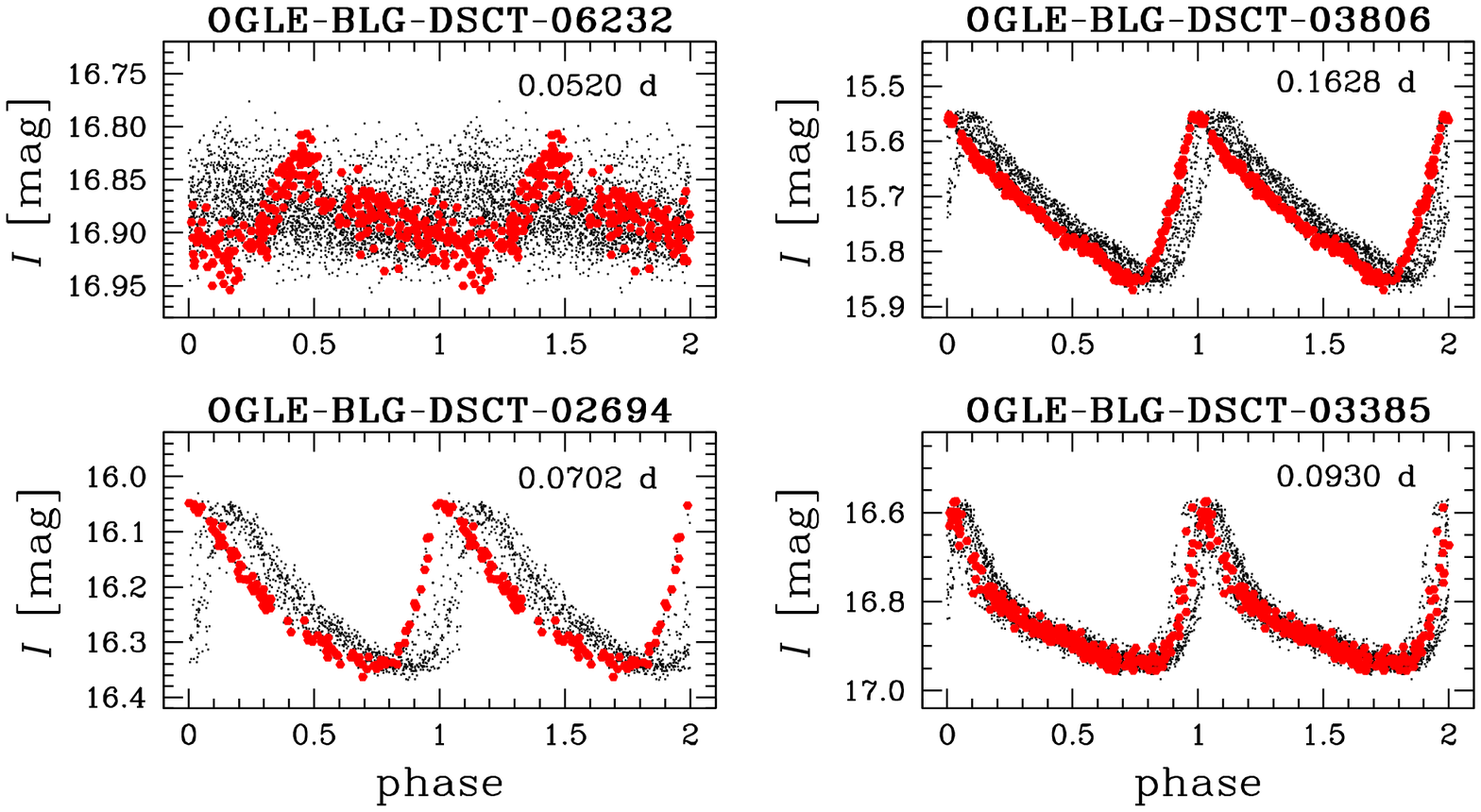}}
\FigCap{Selected $\delta$~Sct stars with evident period changes. Red
points are measurements from 2014 against decade-long data in black.}
\end{figure}

We found dozens of $\delta$~Sct stars with amplitude changes.
Five most prominent examples are presented in Fig.~17. In each case,
we selected two seasons for comparison of the phased light curves.
In some stars, such as OGLE-BLG-DSCT-01200, the amplitude variations
may have a cyclic behavior.

\begin{figure}[htb!]
\centerline{\includegraphics[angle=0,width=130mm]{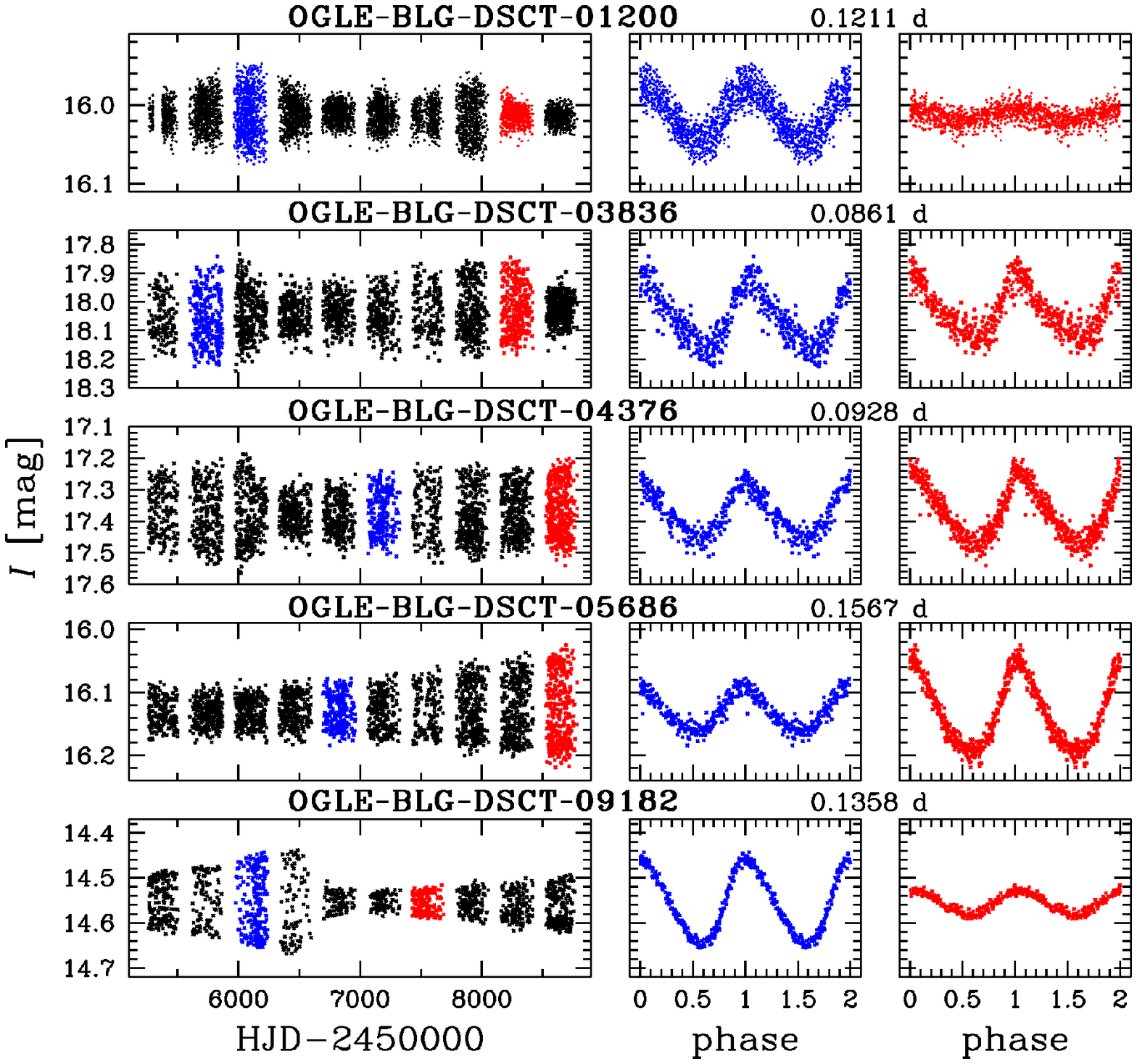}}
\FigCap{Examples of $\delta$~Sct variables with large amplitude
changes. {\it Left column}: time-domain data. {\it Middle and right columns}:
phased light curves from two selected seasons (marked in blue and red
in the time-domain curve). The magnitude range on the vertical axis
is the same for each raw.}
\end{figure}

The correction of the period values applied to the data covering the whole observing
decade led us to the discovery of six $\delta$~Sct stars in which the pulsation
signal greatly weakened or even disappeared. Time-domain light curves of the
stars and power spectra for two selected periods of time are shown in Fig.~18.
Pulsations could stop at all in stars OGLE-BLG-DSCT-00672, OGLE-BLG-DSCT-02036,
OGLE-BLG-DSCT-07325, OGLE-BLG-DSCT-08138. However, a more likely explanation
is that, for a period of time (months to years), the pulsations may have too
low amplitude to be detected in ground-based data. In objects OGLE-BLG-DSCT-05035
and OGLE-BLG-DSCT-05668, the dominant pulsation frequencies are seen in the power
spectra all the time but with changing strength. In the former object, the signal
was strong in years 2010-2011, very weak in years 2014--2015, and then
it became stronger again, as the amplitude increased.

\begin{figure}[htb!]
\centerline{\includegraphics[angle=0,width=130mm]{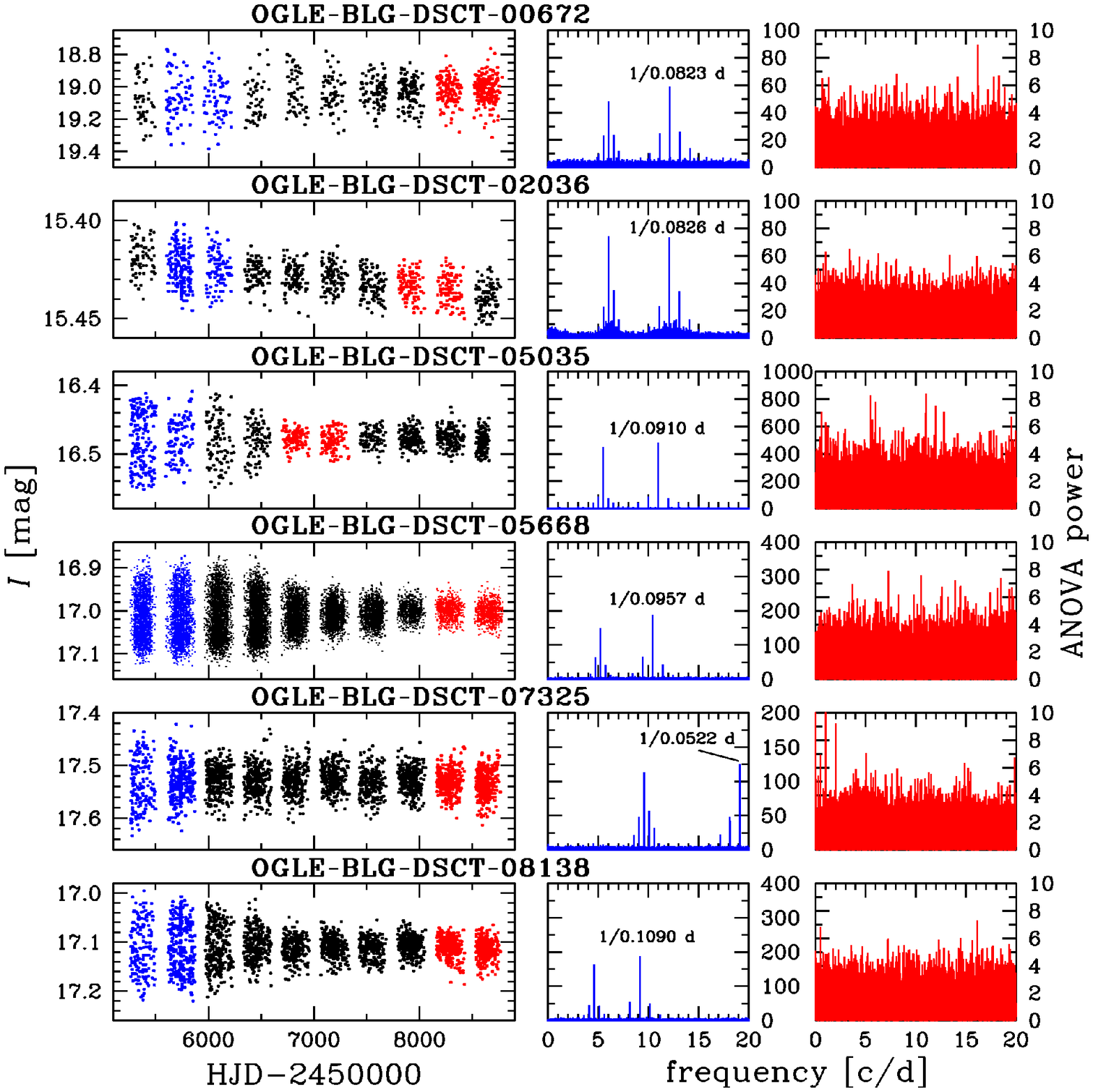}}
\FigCap{$\delta$~Sct variables with significant attenuation or complete
lack of the pulsation signal over a period of time. {\it Left column}:
time-domain data. {\it Middle column}: power spectra from observational
seasons with strong pulsation signal (marked in blue). {\it Right column}:
power spectra from seasons with weak or vanished pulsation signal
(marked in red). The power range in quiescence is magnified.}
\end{figure}

The visual inspection of original light curves of all $\delta$~Sct variables
and light curves of multi-periodic variables after prewhitening for the dominant
period allowed us to find fourteen sources with additional eclipsing or
ellipsoidal binary variations. Table~2 lists parameters of the sources.
Three of the variables were already classified as binary stars in
Soszy\'nski \etal (2016). In Fig.~19, we decompose the signal into the
pulsation and orbital components for four selected objects. In some of the
sources, in particular those with short-period orbital variations, the light
from the $\delta$~Sct star can be blended with the binary signal from other
object in the line of sight. Follow-up spectroscopic observations would confirm
whether the listed $\delta$~Sct stars form physical systems.

\begin{figure}[htb!]
\centerline{\includegraphics[angle=0,width=130mm]{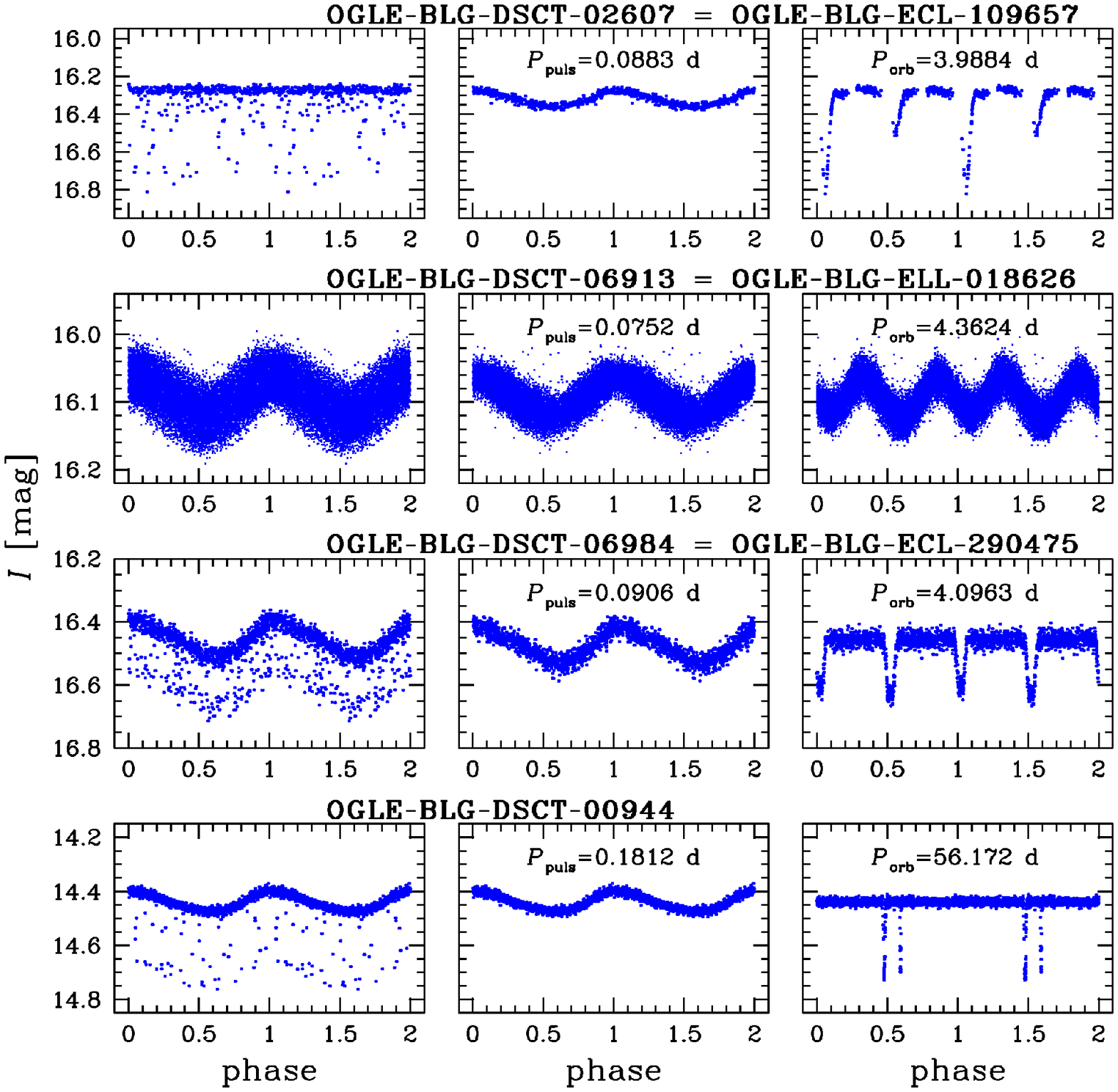}}
\FigCap{Four $\delta$~Sct pulsators with additional binary modulation.
{\it Left column}: original light curves folded with the pulsation period.
{\it Middle and right columns}: pulsation and binary components folded
with the pulsation and orbital periods, respectively. Top three variables
were already classified as binary systems in Soszy\'nski \etal (2016).}
\end{figure}

\begin{table}[htb!]
\centering \caption{\small $\delta$~Sct variables with additional eclipsing
or ellipsoidal modulation}
\medskip
{\small
\begin{tabular}{cccccc}
\hline
Variable  & $\langle I\rangle$ & $\langle V\rangle$ & $P_{\rm puls}$ &  Binarity  & $P_{\rm orb}$    \\
OGLE-BLG- &                    &                    &                &    type    &                 \\
-DSCT-    &        [mag]       &       [mag]        &       [d]      &            &     [d]          \\
\hline
00944     &        14.40       &       16.86        & 0.18121235(5)  &     EA     &  56.172(1)       \\
02533     &        15.67       &       17.23        & 0.08915221(4)  &     EA     &   7.1610(3)      \\
02607     &        16.31       &       17.46        & 0.08832817(4)  &     EB     &   3.98843(4)     \\
03141     &        17.90       &       20.61        & 0.15680356(8)  &     EW     &   0.3390154(2)   \\
04442     &        16.79       &       17.92        & 0.07464503(3)  &     EW     &   4.20527(3)     \\
04527     &        17.44       &       19.56        & 0.06478072(4)  &     EW     &   3.674495(19)   \\
04548     &        16.90       &       18.16        & 0.07158575(1)  &     EW     &   3.15517(5)     \\
05241     &        17.99       &       19.58        & 0.07006432(4)  &     EW     &   0.27519460(12) \\
05566     &        18.05       &        ---         & 0.06349199(1)  &     EW     &   0.2655016(2)   \\
06021     &        16.78       &       17.87        & 0.08031543(4)  &     Ell    &   5.10522(7)     \\
06913     &        16.09       &       17.33        & 0.07524575(5)  &     Ell    &   4.362363(12)   \\
06984     &        16.48       &        ---         & 0.09059884(5)  &     EA     &   4.096252(9)    \\
07078     &        17.51       &       18.77        & 0.04974273(1)  &     EA     &   1.205674(2)    \\
07460     &        17.30       &       18.26        & 0.04894781(1)  &     EW     &   0.4543568(2)   \\
\hline
\end{tabular}}\\
\smallskip
{\footnotesize Mean brightness, pulsation period, detected binarity type,
and orbital period are\\ provided for each variable.}
\end{table}

%%%%%%%%%%%%%%%%%%%%%%%%%%%%%%%%%%%%%%%%%%%%%%%%%%%%%%%%%%%%%%%%%%%%%

\Section{Summary}

Our search for short-period objects in the OGLE-IV photometric data
covering 172 deg$^2$ of the Galactic bulge and spanning the decade
2010--2019 brought the identification of 10~111 genuine $\delta$~Sct-type
variable stars. About 97.3\% of the sample (exactly 9835 stars)
are newly discovered variables. Most of the stars (71.5\%) are multi-mode
pulsators. We showed graphically a huge variety of light curve shapes
of $\delta$~Sct stars, a much greater diverse than observed in other
classical pulsators (\ie Cepheids and RR Lyr stars). The shapes,
periods, and amplitudes in $\delta$~Sct stars vary greatly.
We found six stars in which the pulsation signal had vanished. In one
case, it reappeared. Most likely, the signal was, or in some stars still
is, too faint to be detected in ground-based photometry, obtained even
from a superb astronomical site such as the Las Campanas Observatory, Chile.
There is additional eclipsing or ellipsoidal modulation in fourteen
$\delta$~Sct stars. Spectroscopic follow-up observations would confirm
whether the pulsators possess physically bound companions
and would allow for accurate determination of binary system parameters.

Our sample includes SX Phe-type stars, but it is impossible to
distinguish between the Population II stars and Population I stars
based on the OGLE photometry only. The two groups cannot be separated
by the pulsation period what we hoped for before.

The presented collection of $\delta$~Sct stars has many potential
applications to various areas of stellar astrophysics, such as
asteroseismology, galactic population studies, and distance scale
determination.

%%%%%%%%%%%%%%%%%%%%%%%%%%%%%%%%%%%%%%%%%%%%%%%%%%%%%%%%%%%%%%%%%%%%%

\Acknow{
We thank OGLE observers for their contribution to the collection of
the photometric data over the years. The OGLE project has received
funding from the National Science Centre, Poland, through grant number
MAESTRO 2014/14/A/ST9/00121 to AU. This work has also been supported
by the National Science Centre, Poland, grants OPUS 2016/23/B/ST9/00655
to PP and MAESTRO 2016/22/A/ST9/00009 to IS. HN acknowledges
support from the Polish Ministry of Science and Higher Education under
grant 0192/DIA/2016/45 within the Diamond Grant Programme for years
2016-2021 and Foundation for Polish Science (FNP).}

%%%%%%%%%%%%%%%%%%%%%%%%%%%%%%%%%%%%%%%%%%%%%%%%%%%%%%%%%%%%%%%%%%%%%


\begin{references}

\refitem{Alard, C., and Lupton, R.H.}{1998}{\ApJ}{503}{325}
\refitem{Alcock, C. \etal}{2000}{\ApJ}{536}{798}
\refitem{Balona, L.A., and Dziembowski, W.A.}{2011}{\MNRAS}{417}{591}
\refitem{Bellm, E.C. \etal}{2019}{\PASP}{131}{018002}
\refitem{Blanco, B. M.}{1984}{\AJ}{89}{1836}
\refitem{Breger, M. \etal}{2000}{``Delta Scuti and Related Stars'', ASP Conf. Ser.}{210}{3}
\refitem{Campbell, W.W., and Wright, W.H.}{1900}{\ApJ}{12}{254}
\refitem{Chang, S.-W., Protopapas, P., Kim, D.-W., and Byun, Y.-I.}{2013}{\AJ}{145}{132}
\refitem{Chen, X., Wang, S., Deng, L., de Grijs, R., Yang, M., and Tian, Hao}{2020}{\ApJS}{249}{18}
\refitem{Clement, C.M. \etal}{2001}{\AJ}{122}{2587}
\refitem{Colacevich, A.}{1935}{\PASP}{47}{231}
\refitem{Eggen, O.J.}{1956a}{\PASP}{68}{238}
\refitem{Eggen, O.J.}{1956b}{\PASP}{68}{541}
\refitem{Fath, E.A.}{1935}{\PASP}{47}{232}
\refitem{Fath, E.A.}{1937}{Lick Obs. Bull.}{18}{77}
\refitem{Gaposchkin, S.I.}{1955}{Per. Zvez.}{10}{337}
\refitem{Hamanowicz, A. \etal}{2016}{\Acta}{66}{197}
\refitem{Harris, W.E.}{1996}{\AJ}{112}{1487}
\refitem{Jayasinghe, T. \etal}{2020}{\MNRAS}{493}{4186}
\refitem{Kaluzny, J., Kubiak, M., Szymanski, M., Udalski, A., Krzeminski, W., and Mateo, M.}{1996}{\AAS}{120}{139}
\refitem{Kaluzny, J., and Thompson, I.B.}{2001}{\AA}{373}{899}
\refitem{Kochanek, C.S., \etal}{2017}{\PASP}{129}{104502}
\refitem{Lindblad, P.O., and Eggen, O.J.}{1953}{\PASP}{65}{291}
\refitem{Mazur, B., Krzemi\'nski, W., and Thompson, I.B.}{2003}{\MNRAS}{340}{1205}
\refitem{McNamara, D.H.}{2011}{\AJ}{142}{110}
\refitem{Murphy, S.J., Hey, D., Van Reeth, T., and Bedding, T.R.}{2019}{\MNRAS}{485}{2380}
\refitem{Pietrukowicz, P. \etal}{2013}{\Acta}{63}{379}
\refitem{Pietrukowicz, P. \etal}{2015}{\ApJ}{811}{113}
\refitem{Pietrukowicz, P. \etal}{2017}{Nature Astronomy}{1E}{166}
\refitem{Pigulski, A., Ko{\l}aczkowski, Z., Ramza, T., and Narwid, A.}{2006}{Memorie della Societ\'a Astron. Italiana}{77}{223}
\refitem{Pojma\'nski, G.}{2002}{\Acta}{52}{397}
\refitem{Poleski, R. \etal}{2010}{\Acta}{60}{1}
\refitem{Pych, W., Kaluzny, J., Krzeminski, W., Schwarzenberg-Czerny, A., and Thompson, I.B.}{2001}{\AA}{367}{148}
\refitem{Rodr\'iguez, E., L\'opez-Gonz\'alez, M.J., and L\'opez de Coca, P.}{2000}{\AAS}{144}{469}
\refitem{Rodr\'iguez, E., and L\'opez-Gonz\'alez, M.J.}{2000}{\AA}{359}{597}
\refitem{Rozyczka, M., Thompson, I.B., Pych, W., Narloch, W., Poleski, R., and Schwarzenberg-Czerny, A.}{2017}{\Acta}{67}{203}
\refitem{Samus, N.N., Kazarovets, E.V., Durlevich, O.V., Kireeva, N.N., and Pastukhova, E.N.}{2017}{Astronomy Reports}{61}{80}
\refitem{Schwarzenberg-Czerny, A.}{1996}{\ApJ}{460}{L107}
\refitem{Shappee, B.J., \etal}{2014}{\ApJ}{788}{48}
\refitem{Sollima, A., Cacciari, C., Bellazzini, M., and Colucci, S.}{2010}{\MNRAS}{406}{329}
\refitem{Soszy\'nski, I. \etal}{2010}{\Acta}{60}{17}
\refitem{Soszy\'nski, I. \etal}{2014}{\Acta}{64}{177}
\refitem{Soszy\'nski, I. \etal}{2016}{\Acta}{66}{405}
\refitem{Sterne, T.E.}{1938}{\ApJ}{87}{133}
\refitem{Templeton, M.R., McNamara, B.J., Guzik, J.A., Bradley, P.A., Cox, A.N., and Middleditch, J.}{1997}{\AJ}{114}{1592}
\refitem{Udalski, A., Kubiak, M., Szymanski, M., Kaluzny, J., Mateo, M., and Krzeminski, W.}{1994}{\Acta}{44}{317}
\refitem{Udalski, A., Szymanski, M., Kaluzny, J., Kubiak, M., Mateo, M., and Krzeminski, W.}{1995a}{\Acta}{45}{1}
\refitem{Udalski, A., Olech, A., Szymanski, M., Kaluzny, J., Kubiak, M., Mateo, M., and Krzeminski, W.}{1995b}{\Acta}{45}{433}
\refitem{Udalski, A., Olech, A., Szymanski, M., Kaluzny, J., Kubiak, M., Krzeminski, W., Mateo, M., and Stanek, K.Z.}{1996}{\Acta}{46}{51}
\refitem{Udalski, A., Olech, A., Szymanski, M., Kaluzny, J., Kubiak, M., Mateo, M., Krzeminski, W., and Stanek, K.Z.}{1997}{\Acta}{47}{1}
\refitem{Udalski, A., Szyma\'nski, M.K., and Szyma\'nski, G.}{2015}{\Acta}{65}{1}
\refitem{Walker, M.F.}{1952}{\PASP}{64}{192}
\refitem{Watson, C.L., Henden, A.A., and Price, A.}{2006}{Soc. Astron. Sci. Annu. Symp.}{25}{47}
\refitem{Wo\'zniak, P.R.}{2000}{\Acta}{50}{421}
\refitem{Zloczewski, K., Kaluzny, J., Rozyczka, M., Krzeminski, W., and Mazur, B.}{2012}{\Acta}{62}{357}

\end{references}
\end{document}